\documentclass[prd,preprint, nofootinbib,showpacs,preprintnumbers,amsmath,amssymb]{revtex4}

\usepackage{feynmf}
\usepackage{bm,graphicx,epsfig,color}
\usepackage{dcolumn}

\allowdisplaybreaks

\def\slash#1{#1\!\!\!/\!\,\,}

\newcommand{\order}{ {\cal O} }

\providecommand{\be}{\begin{equation}}
\providecommand{\ee}{\end{equation}}

\newcommand{\nl}{\nonumber \\ }

\begin{document}

\title{
Effective field theory for large logarithms in radiative corrections to electron proton scattering
}

\author{Richard J. Hill}

\affiliation{
  TRIUMF, 4004 Wesbrook Mall, Vancouver, BC, V6T 2A3 Canada
}

\affiliation{
  Perimeter Institute for Theoretical Physics, Waterloo, ON, N2L 2Y5 Canada
}

\affiliation{
  The University of Chicago, Chicago, Illinois, 60637, USA
}

\date{June 9, 2016}

\begin{abstract}
  Radiative corrections to elastic electron-proton scattering are
  analyzed in effective field theory.  A new factorization formula
  identifies all sources of large logarithms in the limit of large
  momentum transfer, $Q^2\gg m_e^2$.  Explicit matching calculations
  are performed through two-loop order.   A renormalization analysis
  in soft-collinear effective theory is  performed to systematically
  compute and  resum large logarithms.  Implications for the
  extraction of charge radii and other observables from scattering
  data are discussed.  The formalism may be applied to other
  lepton-nucleon scattering and $e^+e^-$ annihilation processes. 
\end{abstract}


\maketitle{}

\begin{fmffile}{fmf_brem}

  \fmfcmd{%
    vardef cross_bar (expr p, len, ang) =
    ((-len/2,0)--(0,0))
    rotated (ang + angle direction length(p) of p) shifted point length(p) of p
    enddef;
    style_def crossed expr p =
    cdraw p;
    ccutdraw cross_bar (p, 2mm, 30); ccutdraw cross_bar (p, 2mm, -30)
    enddef;}
  
  \fmfset{arrow_len}{3mm}
  \fmfset{arrow_ang}{12}
  \fmfset{wiggly_len}{2mm}
  \fmfset{wiggly_slope}{75}
  \fmfset{curly_len}{2mm}

\tableofcontents
  
\section{Introduction}

The 2010 measurement of the muonic hydrogen Lamb shift by the CREMA
collaboration~\cite{Pohl:2010zza} determined a value of the proton
electric charge radius, $r_E$, in serious ($\sim 7\sigma$) conflict
with determinations from electronic hydrogen~\cite{Mohr:2012tt} and
electron-proton
scattering~\cite{Sick:2003gm,Bernauer:2010wm,Zhan:2011ji}.  This
``proton radius puzzle'' has far reaching implications across
particle, nuclear and atomic physics.  Taken at face value, in the
absence of explanations beyond the Standard Model, the muonic hydrogen
measurement necessitates a $\gtrsim 5\sigma$ revision of the
fundamental Rydberg constant, in addition to discarding or revising
the predictions from a large body of previous results in both
electron-proton scattering and hydrogen spectroscopy.  Sources of
systematic error that could be impacting electron-proton scattering
measurements, such as incorrect form factor shape assumptions and
inaccurate radiative corrections,  are also at a numerically important
level to  impact neutrino-nucleus scattering, and hence the extraction
of fundamental neutrino parameters, at current and future experiments.  

A recent analysis of global electron-proton scattering data by the
author with Lee and Arrington~\cite{Lee:2015jqa} obtained $r_E=
0.895(20)\,{\rm fm}$ from the high statistics 2010 Mainz A1
dataset~\cite{Bernauer:2013tpr}, and $r_E=0.916(24)\,{\rm fm}$ from
other world data.   A naive average of these results gives
$r_E=0.904(15)\,{\rm fm}$, significantly larger than the muonic
hydrogen determination $r_E(\mu{\rm H})=0.84087(39)\,{\rm fm}$.  The
analysis of Ref.~\cite{Lee:2015jqa} included a critical examination of 
experimental systematic errors and a rigorous treatment of theoretical
uncertainty associated with form factor
shape~\cite{Hill:2010yb,Epstein:2014zua}.    When applied to the
entire $Q^2$ range of the Mainz dataset,  this treatment reinforces
the anomaly with muonic hydrogen.  However, the analysis also revealed
a significant dependence  of the extracted radius on the $Q^2$ range
of data considered.  As noted in this reference, standard models for
radiative corrections were applied.  These models use a phenomenological
ansatz for treating logarithmically enhanced terms, $\sim \alpha^n
\log^{2n}(Q^2/m_e^2)$, where $\log(Q^2/m_e^2) \approx 15$ for $Q^2 \sim {\rm GeV}^2$.
As shown here, such prescriptions fail to
capture subleading logarithms beginning at order $\alpha^2 \log^3(Q^2/m_e^2)$.

More generally, a variety of conflicting conventions and implicit
scheme choices are present in the literature for Born form factors,
charge radii and radiative corrections.  
In this paper, the quantum field theoretical foundation for unambiguously
defining these observables and quantifying uncertainties due to radiative
corrections is constructed.  A new factorization formula is derived
that identifies all sources of large logarithms.
The relation between
conflicting definitions of the charge radius and related observables in the literature is
clarified.  The formalism
may be applied to a range of problems in
lepton-hadron scattering and $e^+ e^-$ annihilation. 
The effective theory analysis simplifies and extends diagrammatic arguments for
the cancellation and exponentiation of infrared singularities in QED~\cite{Yennie:1961ad}. 

The remainder of the paper is structured as follows.   Section~\ref{sec:heavy}
analyzes the scattering problem when particle energies and masses are of comparable
size.  This analysis introduces the soft function that will apply identically
to the more complicated relativistic case.  Section~\ref{sec:rel} considers the
relativistic case where new large logarithms appear.  This analysis
proceeds in stages, considering first the static limit of infinite target mass,
then successively including recoil, structure, and nuclear charge corrections.
The concluding Section~\ref{sec:discuss} summarizes the main results, 
discusses applications, and indicates directions for future work.
Appendix~\ref{sec:RG} lists renormalization constants and conventions employed in the
paper.  Appendix~\ref{sec:convent} compares our preferred Born form factor
convention to others in the literature.
Appendix~\ref{sec:phase} lists relevant phase space integrals.  
Appendix~\ref{sec:2loopfull} gives
details of the computation of two-loop mixed real-virtual corrections in the
static source limit.  Appendix~\ref{sec:2loopeff} presents the same computation
using momentum regions analysis.  

\section{Heavy particle \label{sec:heavy}}

Consider the scattering of a fermion of mass $M$ from a gauge source,
in the regime of energy and momentum transfer $E \sim Q \sim M$, and
including the effects of soft radiation of energy $\Delta E \ll M$.  We will
develop formalism that applies equally well to composite and
elementary particles.  For definiteness in the discussion we refer to
the heavy  particle as a ``proton''.%
\footnote{
  To orient the reader: for the application to electron-proton scattering, 
  the analysis of Section~\ref{sec:heavy} can be viewed as describing
  the ``lower vertex'' (i.e., the proton) in single photon exchange approximation.
  Section~\ref{sec:rel} describes the ``upper vertex'' (i.e., the electron), before
  assembling both pieces and accounting for multiple photon exchange. 
}

The
effective field theory separates physics at the hard scale, with
particle virtualities
$p^2 \sim M^2$,  from physics at the soft scale, $p^2 \sim (\Delta
E)^2$, and enables the resummation of large logarithms, $\log(M/\Delta
E) \gg 1$ using renormalization group methods.  We give a
field-theoretic justification for the conventional separation between
on-shell and Born form factors~\cite{Lee:2015jqa}.  At the same time,
we introduce formalism and notation that will carry over to the more
complicated case of relativistic electron scattering (i.e., $Q^2\gg m^2$)
considered later. 

\begin{figure}[tb]
  \begin{center}
    \parbox{30mm}{
      \begin{fmfgraph*}(80,50)
        \fmfset{arrow_len}{3mm}\fmfset{arrow_ang}{12}
        \fmfleftn{l}{3}
        \fmfrightn{r}{3}
        \fmfbottomn{b}{1}
        \fmf{phantom}{l3,v,r3}
        \fmf{photon, tension=0.6}{b1,v}
        \fmffreeze
        \fmf{fermion}{l3,v,r3}
        \fmfv{d.sh=circle,d.f=empty,d.si=.1w,l=$\times$,label.dist=0}{b1}
        \fmfblob{0.25w}{v}
        \fmflabel{$p$}{l3}
       \fmflabel{$p^\prime$}{r3}
      \end{fmfgraph*}
    }
    \caption{Scattering of proton from electromagnetic source. 
    \label{fig:proton}
    }
  \end{center}
\end{figure}
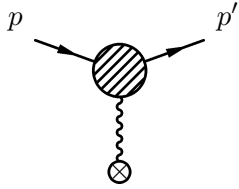

\subsection{Effective theory}

For the process depicted in Fig.~\ref{fig:proton},
introduce timelike unit vectors $v^\mu$ and $v^{\prime\mu}$
via
\begin{align}
  p^\mu=M v^\mu \,, \quad  p^{\prime\mu} = M v^{\prime\mu} \,. 
\end{align}
At factorization scale $\mu\sim M$, hard momentum modes are integrated out, leaving
a low energy effective theory consisting of heavy particle source fields
interacting with soft photons.  
The QED current is matched to an expansion in effective operators, 
\be\label{eq:effops}
J^\mu = \bar{\psi}\gamma^\mu \psi \to \sum_i c_i(\mu, v\cdot v^\prime) \bar{h}_{v^\prime} \Gamma^\mu_i h_v \,, 
\ee
where $h_v$, $h_{v^\prime}$ denote heavy fermion fields satisfying $\slash{v} h_v = h_v$.%
\footnote{For reviews of heavy particle effective theories in the context of QCD and heavy quarks,
  see Refs.~\cite{Neubert:1993mb,Manohar:2000dt}.  NRQED was introduced in Ref.~\cite{Caswell:1985ui}.
  For a discussion of general heavy
  particle effective theories see Ref.~\cite{Heinonen:2012km}.}
The heavy fermion fields interact with soft photons, as described by the effective theory Lagrangian 
\begin{align}\label{eq:Lsoft}
  {\cal L}_{\rm eff.}  = -\frac14 (F^{\mu\nu})^2
  +  \bar{h}_{v} ( iv\cdot \partial + Z e v\cdot A) h_v
  +  \bar{h}_{v^\prime} ( iv^\prime \cdot \partial + Z e v^\prime \cdot A) h_{v^\prime} + \order(1/M) \,,
\end{align}
where $Z=+1$ for the proton,
$A^\mu$ is the electromagnetic field and $F_{\mu\nu} = \partial_\mu A_\nu - \partial_\nu A_\mu$. 

\subsection{One loop matching}

An explicit basis of operator structures in Eq.~(\ref{eq:effops})
respecting the discrete symmetries of the electromagnetic current is
\begin{align}
  \Gamma_1^\mu=\gamma^\mu,  \quad \Gamma_2^\mu = v^\mu + v^{\prime\mu} \,.
\end{align}
For an elementary particle, the matching may be performed
perturbatively.   In the $\overline{\rm MS}$ scheme at renormalization
scale $\mu$, the matching coefficients are~\cite{Neubert:1992tg}
\begin{align}\label{eq:heavy_match}
  c_1(\mu,w) &= 1 - {Z^2 \bar{\alpha} \over 2\pi} \bigg[ ( w f(w) - 1 ) \log{M^2\over \mu^2} - F(w) \bigg] \,,
  \nl
  c_2(\mu,w) &= -{Z^2 \bar{\alpha} \over 4\pi} f(w) \,,
\end{align}
where $w\equiv v\cdot v^\prime$,
\begin{align}\label{eq:rw}
  f(w) &= {1\over \sqrt{w^2-1} } \log(w_+) \,,
  \nl
  F(w) &= {w\over\sqrt{w^2-1}} \bigg[ 2{\rm Li}_2(-w_-) + {\pi^2\over 6} 
     + \frac12\log^2(w_+) - \log(w_+)\log[2(w+1)] + \frac32\log(w_+) \bigg]
  \nl
  &\quad 
  + \frac32 f(w) - 2 \,, 
\end{align}
and for a general quantity $a>1$ we define
\begin{align}\label{eq:apm}
a_\pm \equiv a \pm \sqrt{a^2-1} \,. 
\end{align}
The quantity $\bar{\alpha}$ denotes the running coupling in the
$\overline{\rm MS}$ scheme, $\bar{\alpha}\equiv \alpha(\mu)$.  

The eikonal, $v\cdot A$, nature of the photon coupling in
Eq.~(\ref{eq:Lsoft}) implies that the soft photon matrix element is
universal to the different operator structures $\Gamma_i$ in
Eq.~(\ref{eq:effops}).  This universality becomes manifest with a
Wilson line field redefinition,
\begin{equation}\label{eq:decouple} 
h_v \to  {\cal S}_v h_v \,, \quad {\cal S}_v(x) = \exp\left[
  iZe\int_{-\infty}^0 \!\! ds\, v\cdot A(x+sv) \right] \,,
\end{equation}
that isolates all photon dynamics in a soft-photon Wilson loop, ${\cal
  S}_{v^\prime}^\dagger {\cal S}_v$.  The contribution of soft photons
to the amplitude for the process depicted in Fig.~\ref{fig:proton} is
independent of whether the particle is composite or elementary.  We
define the universal soft form factor to include appropriate
wavefunction renormalization.  Through one loop order this function
reads, 
\vspace{5mm}
\begin{align}\label{eq:FS}
  & F_S(w,\mu) =
  Z_h \left[
  \parbox{25mm}{
      \begin{fmfgraph*}(70,40)
        \fmfset{arrow_len}{3mm}\fmfset{arrow_ang}{12}
        \fmfleftn{l}{3}
        \fmfrightn{r}{3}
        \fmfbottomn{b}{1}
        \fmf{phantom}{l3,v,r3}
        \fmf{photon, tension=0.6}{b1,v}
        \fmffreeze
        \fmf{double}{l3,v,r3}
        \fmfv{d.sh=circle,d.f=empty,d.si=.1w,l=$\times$,label.dist=0}{b1}
        \fmfdot{v}
      \end{fmfgraph*}
    }
    +
    \parbox{25mm}{
      \begin{fmfgraph*}(70,40)
        \fmfset{arrow_len}{3mm}\fmfset{arrow_ang}{12}
        \fmfleftn{l}{3}
        \fmfrightn{r}{3}
        \fmfbottomn{b}{1}
        \fmf{phantom}{l3,x,v,y,r3}
        \fmf{photon, tension=0.6}{b1,v}
        \fmffreeze
        \fmf{double}{l3,x}
        \fmf{double}{x,v,y}
        \fmf{double}{y,r3}
        \fmf{photon,left}{x,y}
        \fmfv{d.sh=circle,d.f=empty,d.si=.1w,l=$\times$,label.dist=0}{b1}
        \fmfdot{v}
      \end{fmfgraph*}
    }
    \right]
    = 1 - {Z^2 \alpha\over 2\pi} \left[ w f(w) - 1 \right] \log{\mu^2\over \lambda^2} 
    \,,
\end{align}

\vspace{5mm}
\noindent
where $\lambda$ is an infinitesimal photon mass acting as IR regulator, and
$Z_h$ is the onshell wavefunction renormalization constant computed from
the lagrangian (\ref{eq:Lsoft}) (cf. Appendix~\ref{sec:RG}). 
The complete (onshell, renormalized) amplitude for the process in Fig.~\ref{fig:proton}
is conventionally expressed as 
\be
\langle J^\mu \rangle = \bar{u}_{v^\prime} \left[ \tilde{F}_1 \gamma^\mu +
\tilde{F}_2 {i\over 2} \sigma^{\mu\nu}(v^{\prime}_\nu-v_\nu) \right] u_v \,,
\ee
where $u_v = u(p)$ is a Dirac spinor and
the onshell  Dirac and Pauli form factors are
\begin{align}\label{eq:ciborn}
  \tilde{F}_1(q^2) &= [c_1(w,\mu)+2c_2(w,\mu) ] F_S(w,\mu) \,,
  \nl
  \tilde{F}_2(q^2) &= - 2c_2(w,\mu) F_S(w,\mu) \,, 
\end{align}
with $q^2 = -2M^2(w-1)$.  
For a strongly interacting composite particle like the proton, perturbative matching is not possible.
In this case, the Wilson coefficients $c_i(w,\mu)$ in Eq.~(\ref{eq:ciborn})
are identified as infrared finite ``Born'' form factors, to be extracted experimentally:
\begin{align}\label{eq:born}
  F_i(q^2)^{\rm Born} &\equiv \tilde{F}_i(q^2) F_S^{-1}(w,\mu=M) \,,
 \end{align}
where the choice $\mu=M$ is part of the Born convention.  
For a discussion of Born form factor extraction from experimental data, see
Ref.~\cite{Lee:2015jqa}.  A comparison to other conventions in the literature
for Born form factors is given in Appendix~\ref{sec:convent}. 

\subsection{Resummation}

To define an infrared finite observable, consider the process depicted in
Fig.~\ref{fig:proton}: scattering of a proton from an electromagnetic
source, allowing radiation of energy $\Delta E \ll M$.  
Suppressing a kinematic prefactor, the cross section
is governed by the factorization formula, 
\begin{align} \label{eq:fac_heavy}
  d\sigma \propto  H\left( {M\over \mu} , v\cdot v^\prime \right)
  S\left( {\Delta E\over \mu}, v\cdot v^\prime, v^0, v^{\prime 0} \right) \,. 
\end{align}
The hard function is
\begin{align}\label{eq:heavy_hard}
  H = \sum_{i,j} c_i(\mu) c_j^*(\mu)
  {\rm Tr}\left( \Gamma_i {1+\slash{v}\over 2} \overline{\Gamma}_j {1+\slash{v}^\prime\over 2} \right) \,. 
\end{align}
The soft function may be expanded according to photon number,
\be
S = S_{0\gamma} + S_{1\gamma} + S_{2\gamma} + \dots \,, 
\ee
and for each contribution we may expand as a series in $\alpha$,
\be
S_{n\gamma} = \sum_{i=n}^\infty \left(\bar{\alpha}\over 4 \pi\right)^i S_{n\gamma}^{(i)} \,. 
\ee
Neglecting real photon emission, 
\begin{align}
S_{0\gamma} &= S(\Delta E=0) = |F_S|^2 \,, 
\end{align}
where $F_S$ is the universal soft form factor, whose one-loop expansion is given in Eq.~(\ref{eq:FS}). 
From the Feynman rules of the lagrangian (\ref{eq:Lsoft}),
the first order real photon correction is
\begin{align}\label{eq:soft1}
S_{1\gamma}^{(1)} &= -(4\pi Z)^2 \int_{\ell^0 \le \Delta E} {d^3\ell \over (2\pi)^3}{1\over 2\ell^0}
\left( {v^\mu \over v\cdot \ell} - {v^{\prime \mu} \over v^\prime \cdot \ell} \right)^2
\nl
&= 4 Z^2 \bigg\{
2\log\left(2\Delta E \over \lambda\right) [ w f(w) - 1 ] + G(w,v^0,v^{\prime 0} ) \bigg\} \,,
\end{align}
where $\ell^0 = \sqrt{\vec{l}^2 + \lambda^2}$, and 
\begin{align}
  G(w,v^0,v^{\prime 0})
  &=
  {v^0 \over \sqrt{ (v^0)^2-1} } \log{v_+^0} + {v^{\prime 0} \over\sqrt{ (v^{\prime 0})^2 -1}} \log{v_+^{\prime 0}}
    + 
  {w\over \sqrt{w^2-1}} \bigg[ \log^2(v^0_+) -\log^2(v^{\prime 0}_+)
    \nl
    &\quad
    + {\rm Li}_2\left( 1 - { v^0_+ \over \sqrt{w^2-1}} ( w_+ v^0 - v^{\prime 0} ) \right)
+ {\rm Li}_2\left( 1 - { v^0_- \over \sqrt{w^2-1}} ( w_+ v^0 - v^{\prime 0} ) \right)
\nl
&\quad
- {\rm Li}_2\left( 1 - { v^{\prime 0}_+ \over \sqrt{w^2-1}} ( v^0 - w_- v^{\prime 0} ) \right)
- {\rm Li}_2\left( 1 - { v^{\prime 0}_- \over \sqrt{w^2-1}} ( v^0 - w_- v^{\prime 0} ) \right)
\bigg]
\,. 
\end{align}
The quantities $v^0_\pm$, $v^{\prime 0}_\pm$, $w_\pm$ are defined by Eq.~(\ref{eq:apm}). 
The total first order correction is thus 
\begin{align}\label{eq:heavy_soft} 
  S^{(1)} = Z^2 \bigg\{ 
  8 \log\left( {2\Delta E \over \mu}\right) [ wf(w) - 1 ]
  + 4 G(w,v^0,v^{\prime 0}) \bigg\} \,.
\end{align}
When $\Delta E \ll M$, large logarithms are present regardless
of the choice for factorization scale $\mu$ in Eq.~(\ref{eq:fac_heavy}). 
This is seen explicitly in the one-loop corrections for the hard function
in Eqs.~(\ref{eq:heavy_match}) and (\ref{eq:heavy_hard}),
and for the soft function in Eq.~(\ref{eq:heavy_soft}). 
The following renormalization analysis systematically resums large logarithms
to all orders in perturbation theory.

The anomalous dimension of the effective
operators (\ref{eq:effops}) relates the renormalization of the hard function
to the cusp anomalous dimension for QED~\cite{Korchemsky:1987wg,Kilian:1992tj},  (cf. Appendix~\ref{sec:RG})
\be\label{eq:hanom}
{d\over d\log\mu} H(\mu) = 2\Gamma_{\rm cusp}(w) H(\mu) \,. 
\ee
Expanding in $\alpha$, 
\begin{align}\label{eq:Gcusp}
\Gamma_{\rm cusp}(w) &= \sum_{n=0}^\infty \left(\bar{\alpha}\over 4\pi\right)^{n+1} \Gamma^{\rm cusp}_n(w) \,,
\end{align}
where the leading terms are (cf. Appendix~\ref{sec:RG})
\begin{align}\label{eq:Gi}
  \Gamma^{\rm cusp}_0(w) = 4 [ wf(w) - 1 ]
  \,, \quad
\Gamma^{\rm cusp}_1(w) = -\frac{20}{9} n_f \Gamma^{\rm cusp}_0 \,. 
\end{align}
Here $n_f$ denotes the number of light fermions in the effective theory.
In this example, we take the muon mass, proton mass and
other hadronic scales as large compared to $\Delta E$,
and work with $n_f=1$ in the regime with formal power counting
$m=m_e \sim \Delta E \ll E \sim m_\mu \sim m_p=M$.%
\footnote{It is straightforward to include perturbative corrections due to the muon.
  }
Solution of Eq.~(\ref{eq:hanom}) then yields
\begin{align}\label{eq:soft_resum}
{H(\mu_H)\over H(\mu_L)} &= {S(\mu_L)\over S(\mu_H)} 
\nl
&
= \exp\bigg\{ -{\Gamma^{\rm cusp}_0(w) \over \beta_0} \bigg[
  \log{\alpha(\mu_H)\over\alpha(\mu_L)}
  + {1\over 4\pi}\left({\Gamma^{\rm cusp}_1\over \Gamma^{\rm cusp}_0} - {\beta_1\over\beta_0} \right)
  \big( \alpha(\mu_H) - \alpha(\mu_L) \big)  + \dots \bigg] \bigg\} 
\nl
&= \exp\bigg\{ 4[ wf(w) -1 ]\bigg[ {\alpha\over 4\pi}\log{\mu_H^2\over \mu_L^2}
  + \left(\alpha\over 4\pi\right)^2
  \bigg(
   \frac23 \log^2{\mu_H^2 \over \mu_L^2}
    + \frac43 \log{\mu_H^2\over \mu_L^2}\log{\mu_L^2\over m^2}
    \nl
    &\quad
    - {20\over 9} \log{\mu_H^2\over\mu_L^2}
    \bigg)
    + \order(\alpha^3)
  \bigg]
  \bigg\} \,,
\end{align}
where the result in the last line is expressed in terms of the low energy, onshell,
fine structure constant $\alpha$.  

To connect with observables such as the Born form factors  (\ref{eq:born}) defined
at $\mu \sim M$, we may expand soft functions in perturbation theory
at the scale $\mu_L \sim \Delta E \sim m$, where no large logarithms appear,
as in Eq.~(\ref{eq:heavy_soft}). 
We may then use Eq.~(\ref{eq:soft_resum}) to evaluate the soft function 
appearing in Eq.~(\ref{eq:born}) at $\mu_H \sim M$, 
systematically controlling large logarithms. 

We remark that a simple exponentiation ansatz,
\begin{align}
S \to \exp\left[ {\alpha\over 4\pi} S^{(1)} \right] \,,
\end{align}
fails to capture logarithmically enhanced terms beginning at order
$\alpha^2\log^2[ M^2 / (\Delta E)^2]$.  Such terms are below typical
experimental accuracies for $w=\order(1)$.  However, at large recoil, $w\gg 1$, 
additional factors involving large logarithms, $\log(w)$, appear.  We
turn now to this case, where control of logarithmically enhanced
corrections beyond first order in $\alpha$ is essential.  

\section{Relativistic particle \label{sec:rel}} 

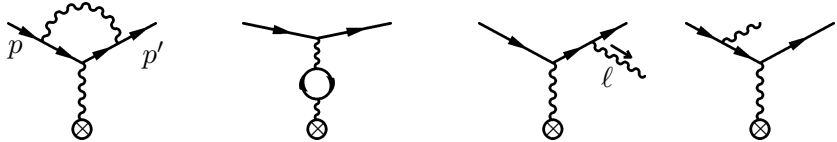
\begin{figure}[tb]
  \begin{center}
    \parbox{30mm}{
      \begin{fmfgraph*}(70,40)
        \fmfset{arrow_len}{3mm}\fmfset{arrow_ang}{12}
        \fmfleftn{l}{3}
        \fmfrightn{r}{3}
        \fmfbottomn{b}{1}
        \fmf{phantom}{l3,x,v,y,r3}
        \fmf{photon, tension=0.6}{b1,v}
        \fmffreeze
        \fmf{fermion,label=$p$, label.dist=.05w}{l3,x}
        \fmf{fermion}{x,v,y}
        \fmf{fermion,label=$p^\prime$, label.dist=.05w}{y,r3}
        \fmf{photon,left}{x,y}
        \fmfv{d.sh=circle,d.f=empty,d.si=.1w,l=$\times$,label.dist=0}{b1}
      \end{fmfgraph*}
    }
    \parbox{30mm}{
      \begin{fmfgraph*}(70,40)
        \fmfleftn{l}{3}
        \fmfrightn{r}{3}
        \fmfbottomn{b}{1}
        \fmf{fermion}{l3,v,r3}
        \fmf{photon}{b1,x}
        \fmf{photon}{y,v}
        \fmf{fermion,left,tension=0.5}{x,y,x}
        \fmfv{d.sh=circle,d.f=empty,d.si=.1w,l=$\times$,label.dist=0}{b1}
      \end{fmfgraph*}
    }
    \parbox{30mm}{
      \begin{fmfgraph*}(70,40)
        \fmfset{arrow_len}{3mm}\fmfset{arrow_ang}{12}
        \fmfleftn{l}{3}
        \fmfrightn{r}{3}
        \fmfbottomn{b}{1}
        \fmf{phantom}{l3,x,v,y,r3}
        \fmf{photon, tension=0.6}{b1,v}
        \fmffreeze
        \fmf{fermion}{l3,v,y,r3}
        \fmf{photon,label=$\ell$, l.d=0.05w, tag=1}{y,r2}
        \fmfipath{p[]}
        \fmfiset{p1}{vpath1(__y,__r2)}
        \fmfi{crossed}{ point 2length(p1)/5 of p1 shifted (thick*(0,2)) -- point 3length(p1)/4 of p1 shifted (thick*(0,2))}        
        \fmfv{d.sh=circle,d.f=empty,d.si=.1w,l=$\times$,label.dist=0}{b1}
      \end{fmfgraph*}
    }
\hspace{-5mm}    
    \parbox{30mm}{
      \begin{fmfgraph*}(70,40)
        \fmfset{arrow_len}{3mm}\fmfset{arrow_ang}{12}
        \fmfleftn{l}{3}
        \fmfrightn{r}{3}
        \fmfbottomn{b}{1}
        \fmftopn{t}{1}
        \fmf{phantom}{l3,x,v,y,r3}
        \fmf{photon, tension=0.6}{b1,v}
        \fmffreeze
        \fmf{fermion}{l3,x,v,r3}
        \fmf{photon}{x,t1}
        \fmfv{d.sh=circle,d.f=empty,d.si=.1w,l=$\times$,label.dist=0}{b1}
      \end{fmfgraph*}
    }
    \caption{First order radiative corrections to electron scattering from static source. 
    \label{fig:first}
  }
  \end{center}
\end{figure}

When particle velocities satisfy $v\cdot v^\prime \gg 1$, new large
logarithms appear in perturbation theory which are not resummed by the
renormalization analysis in the heavy particle effective theory of the
previous section.  For example, $c_i(\mu,v\cdot v^\prime)$ in Eq.~(\ref{eq:heavy_match})
contains large logarithms, $\log(v\cdot v^\prime)$, regardless of the choice for
factorization scale $\mu$.   
In order to isolate and resum these additional
large logarithms, we must extend the effective theory to include
collinear degrees of freedom~\cite{Bauer:2000ew,Bauer:2000yr,Bauer:2001ct,Bauer:2001yt,Chay:2002vy,Beneke:2002ph,Hill:2002vw,Becher:2014oda}.  Before turning to the effective theory
description, let us examine the explicit two-loop calculation for
relativistic 
electron-proton scattering in the static source limit.  We will
then 
perform the effective theory analysis in this limit before 
including arbitrary recoil corrections,
and radiative corrections involving the proton. 

\subsection{Two loop corrections in static limit}

To isolate the essential points, let us consider the problem of
relativistic unpolarized electron-proton scattering in the
static-source limit of large proton mass:
$m \ll E \ll M$, where $m$ and $M$ denote
the electron and proton masses and $E$ is the electron energy.
Neglecting power corrections in $m/E$, and working to first order
in nuclear charge (i.e., single photon exchange), 
the cross section may be written 
\be\label{eq:raddef}
d\sigma = { (d\sigma)_{\rm Mott}  \over [1-\hat{\Pi}(q^2)]^2 }
\left( 1 + \delta_{e} + \delta_{e\gamma} + \delta_{e\gamma\gamma} + \dots
\right) \,,
\ee
where $(d\sigma/d\Omega)_{\rm Mott} = \alpha^2 \cos^2(\theta/ 2) / [ 4 E^2 \sin^4(\theta/ 2)] $
is the tree-level, Mott, cross section, 
and $\hat{\Pi}(q^2)$ is the photon vacuum polarization function.
Each term $\delta_X$ in Eq.~(\ref{eq:raddef}) corresponds to different numbers of
final state photons and is expanded according to
$\delta_{X} =  \sum_{n=0}^\infty \left(\alpha\over 4 \pi\right)^n \delta_{X}^{(n)}$.

Consider radiative corrections at first order in $\alpha$, cf. Fig.~\ref{fig:first}.  
Regulating infrared divergences with an infinitesimal photon mass $\lambda$,
corrections with just an electron in the final state are 
\be\label{eq:virt}
1 + \delta_e = [F_1(q^2,m^2,\lambda^2)]^2 \,,
\quad
F_1 = 1 + \sum_{n=1}^\infty \left(\alpha\over 4 \pi\right)^n F_1^{(n)} \,,
  \ee
  where $F_1$ is the Dirac form factor of the electron. 
  At large spacelike momentum transfer $Q^2=-q^2\gg m^2$, the limit of
  Eq.~(\ref{eq:ciborn}), using  Eqs.~(\ref{eq:heavy_match}) and (\ref{eq:FS}),
  yields 
[$L\equiv \log(Q^2/m^2)$]
\begin{align}\label{eq:F11}
F_1^{(1)}&= 4 \log{\lambda\over m} \left( L - 1 \right) 
- L^2 + 3 L - 4 + {\pi^2\over 3}  \,. 
\end{align}
Real radiation corrections are given by the limit of Eq.~(\ref{eq:soft1}), 
\begin{align}\label{eq:eg1}
  \delta_{e\gamma}^{(1)} &=
  -8\left( \log{E\over \Delta E} + \log{\lambda\over m} \right) (L-1)
+ 2 L^2
+ 4{\rm Li}_2\left( \cos^2{\theta\over 2} \right) - {4\pi^2\over 3} \,,
\end{align}
where a cut $\ell^0 \le \Delta E \ll E$ is placed on photon energy. 
The total first order correction, $\delta^{(1)}= \delta_{e}^{(1)} + \delta_{e\gamma}^{(1)}$,
is infrared finite. 

\begin{figure}[t]
  \begin{center}
    \parbox{30mm}{
      \begin{fmfgraph*}(80,40)
        \fmfleftn{l}{4}
        \fmfrightn{r}{4}
        \fmfbottomn{b}{1}
        \fmf{phantom}{l4,x1,x2,v,y1,y2,r4}
        \fmf{photon, tension=0.6}{b1,v}
        \fmffreeze
        \fmf{fermion}{l4,x1,v,y1,y2,r4}
        \fmf{phantom,tag=1}{y1,r2}
        \fmfipath{p[]}
        \fmfiset{p1}{vpath1(__y1,__r2)}
        \fmfi{photon}{ point 0length(p1) of p1 -- point length(p1)/2 of p1}
        \fmf{photon,left=0.5}{x1,y2}
        \fmfv{d.sh=circle,d.f=empty,d.si=.1w,l=$\times$,label.dist=0}{b1}
      \end{fmfgraph*}
    }
\hspace{-5mm}    
    \parbox{30mm}{
      \begin{fmfgraph*}(80,40)
        \fmfset{arrow_len}{3mm}\fmfset{arrow_ang}{12}
        \fmfleftn{l}{5}
        \fmfrightn{r}{5}
        \fmfbottomn{b}{1}
        \fmf{phantom}{l5,x1,x2,v,y1,y2,r5}
        \fmf{photon, tension=0.6}{b1,v}
        \fmffreeze
        \fmf{fermion}{l5,x2,v,y1,y2,r5}
        \fmf{phantom,tag=1}{y2,r3}
        \fmfipath{p[]}
        \fmfiset{p1}{vpath1(__y2,__r3)}
        \fmfi{photon}{ point 0length(p1) of p1 -- point 2length(p1)/3 of p1}
        \fmf{photon,left=0.5}{x2,y1}
        \fmfv{d.sh=circle,d.f=empty,d.si=.1w,l=$\times$,label.dist=0}{b1}
      \end{fmfgraph*}
    }
    \hspace{-5mm}
        \parbox{30mm}{
          \begin{fmfgraph*}(80,40)
            \fmfset{arrow_len}{3mm}\fmfset{arrow_ang}{12}
            \fmfleftn{l}{4}
            \fmfrightn{r}{4}
            \fmfbottomn{b}{1}
            \fmf{phantom}{l4,x1,x2,v,y1,y2,y3,r4}
            \fmf{photon, tension=0.6}{b1,v}
            \fmffreeze
            \fmf{fermion}{l4,v,y1,y2,y3,r4}
            \fmf{phantom,tag=1}{y2,r2}
            \fmfipath{p[]}
            \fmfiset{p1}{vpath1(__y2,__r2)}
            \fmfi{photon}{ point 0length(p1) of p1 -- point 2length(p1)/3 of p1}
            \fmf{photon,left}{y1,y3}
            \fmfv{d.sh=circle,d.f=empty,d.si=.1w,l=$\times$,label.dist=0}{b1}
          \end{fmfgraph*}
        }
        \parbox{30mm}{
        \begin{fmfgraph*}(80,40)
          \fmfset{arrow_len}{3mm}\fmfset{arrow_ang}{12} 
           \fmfleftn{l}{5}
            \fmfrightn{r}{5}
            \fmfbottomn{b}{1}
            \fmf{phantom}{l5,x1,x2,v,y1,y2,y3,r5}
            \fmf{photon, tension=0.6}{b1,v}
            \fmffreeze
            \fmf{fermion}{l5,v,y1,y2,y3,r5}
            \fmf{phantom,tag=1}{y3,r3}
            \fmfipath{p[]}
            \fmfiset{p1}{vpath1(__y3,__r3)}
            \fmfi{photon}{ point 0length(p1) of p1 -- point 2length(p1)/3 of p1}
            \fmf{photon,tension=0.5, left=2}{y1,y2}
            \fmfv{d.sh=circle,d.f=empty,d.si=.1w,l=$\times$,label.dist=0}{b1}
          \end{fmfgraph*}
        }
        \parbox{30mm}{
          \begin{fmfgraph*}(70,40)
            \fmfleftn{l}{3}
            \fmfrightn{r}{3}
            \fmfbottomn{b}{1}
            \fmf{phantom}{l3,aa,v,bb,r3}
            \fmf{photon}{b1,x}
            \fmf{photon}{y,v}
            \fmf{fermion,left,tension=0.5}{x,y,x}
            \fmffreeze
            \fmf{fermion}{l3,v,bb,r3}
            \fmf{phantom,tag=1}{bb,r2}
            \fmfipath{p[]}
            \fmfiset{p1}{vpath1(__bb,__r2)}
            \fmfi{photon}{ point 0length(p1) of p1 -- point 2length(p1)/3 of p1}
            \fmfv{d.sh=circle,d.f=empty,d.si=.1w,l=$\times$,label.dist=0}{b1}
          \end{fmfgraph*}
        }
        \caption{Second order radiative corrections to electron scattering from static source. 
          Diagrams involving photon emission from the initial state electron are not shown. 
          \label{fig:second}
        }
  \end{center}
\end{figure}
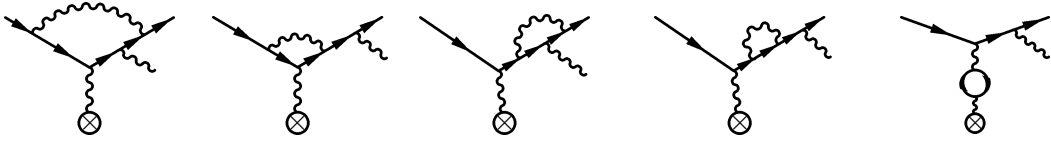

Second order corrections containing two-photon final states (``double
bremsstrahlung'') are
\begin{align}\label{eq:double}
  \delta_{e\gamma\gamma}^{(2)} &= {1\over 2!}
  \int_{\Delta E} {d^3\ell \over \pi\ell^0} 
{d^3\ell^\prime \over \pi\ell^{\prime 0}} 
\left( {Q^2 \over p\cdot\ell p^\prime\cdot\ell } - {m^2\over (p\cdot \ell)^2} - {m^2\over (p^\prime \cdot \ell)^2} 
\right)
\left( {Q^2 \over p\cdot\ell^\prime p^\prime\cdot\ell^\prime }
- {m^2\over (p\cdot \ell^\prime)^2}
- {m^2\over (p^\prime \cdot \ell^\prime)^2} 
\right)
\nl
&= {1\over 2!}\left[ \delta_{e\gamma}^{(1)} \right]^2 - {16\pi^2\over 3} (L-1)^2  \,,
\end{align}
where a cut $\ell^0 + \ell^{\prime 0} \le \Delta E$ is placed on
photon energy. 
Contributions to second order mixed real-virtual corrections are displayed in
Fig.~\ref{fig:second}.  The computation of these contributions is
described in Appendix~\ref{sec:2loopfull}. 
After renormalization, and neglecting power suppressed contributions,
the result takes the simple form 
\begin{align}\label{eq:deltahat}
  \delta_{e\gamma}^{(2)}
    &= \delta_e^{(1)} \delta_{e\gamma}^{(1)}  \,, 
\end{align}
where $\delta_e^{(1)} = 2 F_1^{(1)}$ in Eq.~(\ref{eq:F11}) and $\delta_{e\gamma}^{(1)}$
is given in Eq.~(\ref{eq:eg1}). 
Finally, second order virtual corrections, $\delta^{(2)}_e$, are given by expanding
Eq.~(\ref{eq:virt})~\cite{Burgers:1985qg,Kniehl:1989kz,Mastrolia:2003yz}.
The complete second order correction may be written
\begin{align}\label{eq:tot}
  \delta^{(2)}
  &= {1\over 2!}[\delta^{(1)}]^2
  -{8\over 9}L^3 + \left({76\over 9}-{16\pi^2\over 3}\right)L^2
  + \bigg( -{979\over 27}
  + {52\pi^2\over 9} + 48\zeta(3) \bigg) L
  + {4252\over 27} + {31\pi^2\over 3}
    \nl
    &\qquad
- 16\pi^2\log{2}
    - 72 \zeta(3) - {64\pi^4\over 45}
  \,.
\end{align}

\begin{figure}[tb]
  \begin{center}
    \includegraphics[width=0.6\textwidth]{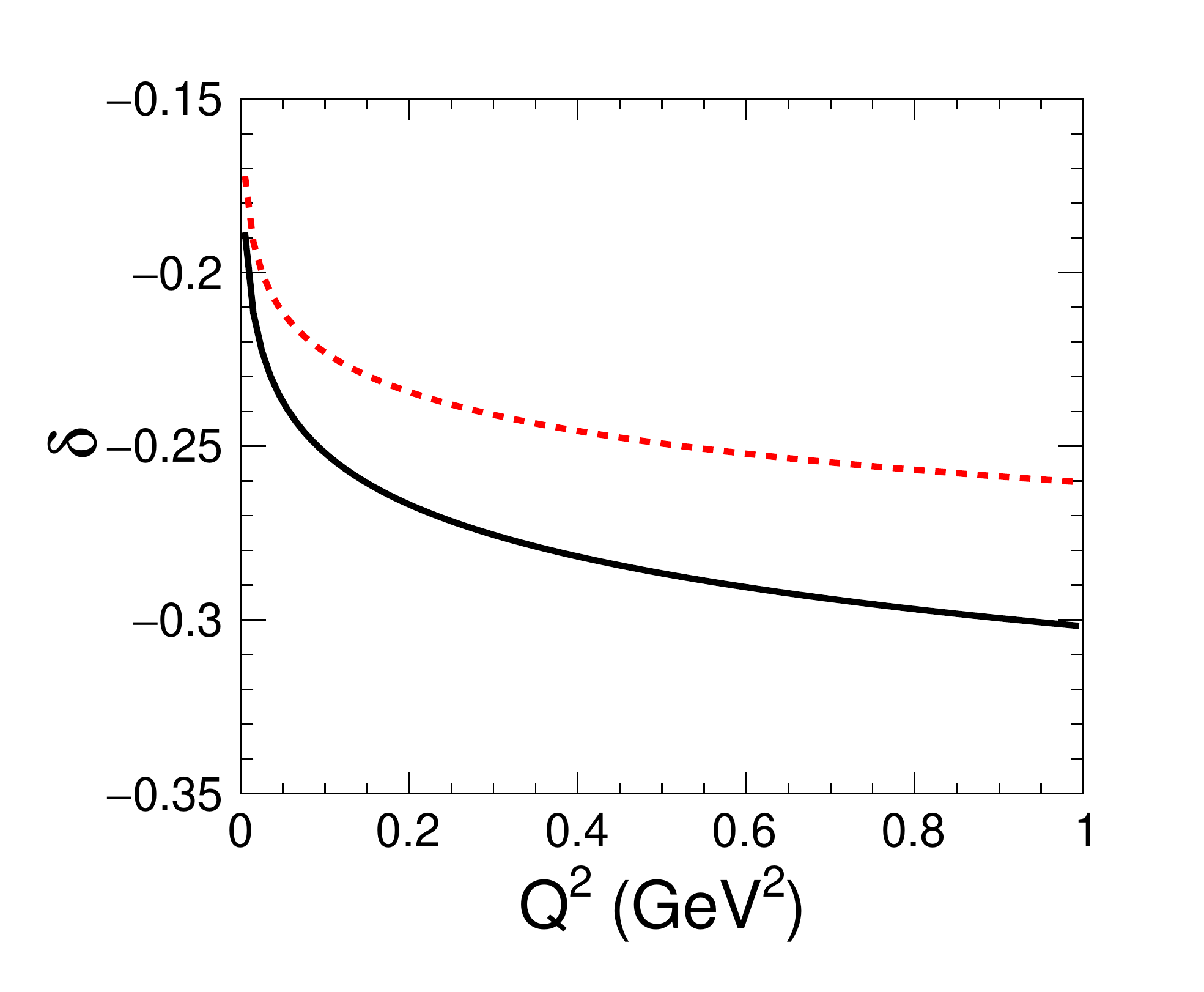}
    \caption{
      Radiative correction $\delta$ in static source limit
      for $E=1\,{\rm GeV}$, $\Delta E= 5\,{\rm MeV}$, computed
      at first (bottom, blue, curve) and second (top, red, curve)
      in $\alpha$. \label{fig:delta}
    }
  \end{center}
\end{figure}

Fig.~\ref{fig:delta} displays the total correction $\delta$ at first and
second order in perturbation theory, for illustrative values
$E=1\,{\rm GeV}$, $\Delta E=5\,{\rm MeV}$.
Logarithmically enhanced
corrections beginning at order $\alpha^2 L^3$ are not captured
by a simple exponentiation ansatz, $\delta \to \exp[{\alpha \over 4 \pi}\delta^{(1)}]$.
In the next section we derive the effective theory that allows identification and
resummation of large logarithms. 

\subsection{Effective theory: matching}

To determine the origin of the different contributions in Eq.~(\ref{eq:tot}),
and to systematically resum large logarithms in perturbation theory, let us
construct an effective theory to separate the physics at different energy
scales.
We focus on the formal counting $m^2 \sim (\Delta E)^2$ and
$Q^2 \gg m^2$ (i.e., $v\cdot v^\prime \gg 1$).
Appendix~\ref{sec:2loopeff} outlines an effective
operator analysis analogous to Eqs.~(\ref{eq:effops}) and (\ref{eq:Lsoft}). 
In place of Eq.~(\ref{eq:fac_heavy}),
the new factorization formula, valid up to $\order(m^2/Q^2)$ corrections
and verified explicitly through two-loop order (cf. Appendices~\ref{sec:2loopfull} and \ref{sec:2loopeff}), reads 
\begin{align}\label{eq:facfull}
  d\sigma \propto H\left( {Q^2\over \mu^2} \right)
  J\left({m^2\over \mu^2} \right)
  R\left( {m^2\over \mu^2}, {p\cdot p^\prime \over m^2} \right)
  S\left( {\Delta E\over \mu}, {p\cdot p^\prime \over m^2}, {E\over m}, {E^\prime \over m} \right) \,.
\end{align}

The explicit matching with QED is most easily performed using dimensional regularization,
where dimensionful but scaleless integrals vanish.
The (bare, unrenormalized) hard function is then [$4\pi\alpha_{\rm bare} \equiv e_{\rm bare}^2 (4\pi)^\epsilon e^{-\gamma_E \epsilon}$]
\begin{align}\label{eq:hard}
  \sqrt{H^{\rm bare}}
  \equiv F_H^{\rm bare}
  = F_1(q^2,m^2=0,\lambda^2=0)
  = 1 + \sum_{i=1}^{\infty} \left( \alpha_{\rm bare} Q^{-2\epsilon} \over 4\pi \right)^i F_{Hi}^{\rm bare}  \,,
\end{align}
where results for $F_1(q^2,0,0)$ through two-loop order are~\cite{Moch:2005id,Gehrmann:2005pd},%
\footnote{ There is a transcription error in the $\order(\epsilon^2)$
  coefficient in Eq.~(15) of Ref.~\cite{Gehrmann:2005pd}:
  $-47\pi^4/2880$ should be replaced by $+47\pi^4/2880$, in accordance
  with Eq.~(17) of the same reference~\cite{gehrmann}.  }
\begin{align}
  F_{H1}^{\rm bare} &= -{2\over \epsilon^2} - {3\over \epsilon} - 8 + \zeta_2
  + \epsilon\left( -16 + {\pi^2\over 4} + {14\over 3} \zeta_3 \right)
  + \epsilon^2\left( -32 + {2\pi^2\over 3} + 7 \zeta_3 + {47\over 720}\pi^4 \right) + \order(\epsilon^3) \,,
  \nl
  F_{H2}^{\rm bare} &= {2\over \epsilon^4} + {6\over \epsilon^3} + {1\over\epsilon^2}\left({41\over 2}-2\zeta_2\right)
  +{1\over\epsilon}\left({221\over 4}-{64\over 3}\zeta_3\right)
  + {1151\over 8} + {17\over 2}\zeta_2 - 58\zeta_3 -13\zeta_2^2
  \nl
  &\quad 
  + 2 n_f\bigg[ {1\over 3\epsilon^3} + {14\over 9\epsilon^2}
    + {1\over \epsilon}
    \left({353\over 54} + {\zeta_2\over 3}\right)
    +{7541\over 324} + {14\zeta_2\over 9} - {26\zeta_3\over 9}
    \bigg]
  +\order(\epsilon) \,.
\end{align}
In the $\overline{\rm MS}$ scheme, we define (at $n_f=1$)
\begin{align}
 F_H(\mu) = Z_H F_H^{\rm bare} \,,
\end{align}
with the renormalization constant,
\begin{align}
  Z_H &= 1 + {\bar{\alpha}\over 4\pi}\bigg[ {2\over \epsilon^2}
    + {1\over\epsilon}\left( -2\log{Q^2\over\mu^2} + 3 \right) \bigg]
  + \left({\bar{\alpha}\over 4\pi}\right)^2\bigg[
    {2\over \epsilon^4} + {1\over\epsilon^3}\left( -4\log{Q^2\over\mu^2} + 8\right)
    \nl
    &\quad
    + {1\over\epsilon^2}\left( 2\log^2{Q^2\over\mu^2} - {22\over 3}\log{Q^2\over\mu^2}
    + {97\over 18}
    \right)
    + {1\over\epsilon}\left( {20\over 9}\log{Q^2\over \mu^2} - {179\over 108}
    -{4\pi^2\over 3} + 12\zeta_3 \right)
    \bigg] + \order(\alpha^3) \,. 
\end{align}
The explicit renormalized hard function is 
\begin{align}\label{eq:Hmu}
F_H(\mu)
  &=
  1 + {\bar{\alpha}\over 4\pi}\bigg[ -\log^2{Q^2\over\mu^2} + 3\log{Q^2\over\mu^2} -8 + {\pi^2\over 6} \bigg]
  \nl &\quad
  + \left(\bar{\alpha}\over 4\pi\right)^2
  \bigg[ \frac12 \log^4{Q^2\over\mu^2} - \frac{31}{9} \log^3{Q^2\over\mu^2}
    + \bigg( {301\over 18}
    - {\pi^2\over 6} \bigg) \log^2{Q^2\over\mu^2}
    \nl &\quad
    + \left( -{2051\over 54} -{35\pi^2\over 18} + 24\zeta_3 \right)\log{Q^2\over\mu^2}
    + {235\pi^2\over 54} - {266\zeta_3\over 9}
    + {36995\over 648} - {83\pi^4\over 360} \bigg]
  + \order(\alpha^3) \,,
\end{align}
where $\bar{\alpha} = \alpha(\mu)$ is the $\overline{\rm MS}$ QED coupling with $n_f=1$ at
renormalization scale $\mu$ (for a summary of renormalization constants and conventions
see Appendix~\ref{sec:RG}).

The soft function in Eq.~(\ref{eq:facfull}) is defined identically to the soft function
in Eq.~(\ref{eq:fac_heavy}), and for virtual corrections becomes trivial ($S=1$) at $\lambda=0$.
The product of the (bare, unrenormalized) jet and 
remainder functions (defined separately below) is thus
\begin{align}\label{eq:R}
  \sqrt{(JR)^{\rm bare}} = F_{JR}^{\rm bare} =
       { F_1(q^2,m^2,\lambda^2=0) \over F_1(q^2,m^2=0,\lambda^2=0) }
  = 1 + \sum_{i=1}^\infty \left( \alpha_{\rm bare} m^{-2\epsilon} \over 4\pi \right)^i F_{JRi}^{\rm bare} 
  \,,
\end{align}
where results for $F_1(q^2,m^2,0)$ through two-loop order are given in
Refs.~\cite{Hoang:1995ex,Bernreuther:2004ih}.
These results imply~\cite{Becher:2007cu}, (now at $n_f=1$)
\begin{align}\label{eq:FJR12}
  F_{JR1}^{\rm bare} &= {2\over \epsilon^2} + {1\over \epsilon} + {\pi^2\over 6} + 4
  + \epsilon\left( 8 + {\pi^2\over 12} - {2\zeta_3\over 3} \right)
  + \epsilon^2\left( 16 - {\zeta_3\over 3} + {\pi^4\over 80} + {\pi^2\over 3} \right) + \order(\epsilon^3) \,,
  \nl
  F_{JR2}^{\rm bare} &= {2\over \epsilon^4} + {4\over 3\epsilon^3}
  + {1\over\epsilon^2}\bigg( {145\over 18}  + {\pi^2\over 3} \bigg)
  + {1\over\epsilon}\bigg( \frac{1405}{108}-{11\pi^2\over 9} + {32\zeta_3\over 3} \bigg)
  + {58957\over 648} + {397\pi^2\over 108} -{62\zeta_3\over 9}
  \nl
  &\quad
  - 8\pi^2\log{2}
  -{77\pi^4\over 180}
  + \log{Q^2\over m^2}\bigg( -{4\over 3\epsilon^2} + {20\over 9\epsilon} - {112\over 27} - {2\pi^2\over 9} \bigg) 
  + \order(\epsilon) \,.
\end{align}
The product $F_H F_{JR}$ represents the matching coefficient onto the soft operator after integrating out
the electron mass scale.   In the $\overline{\rm MS}$ scheme for the $n_f=0$ theory,
we write
\begin{align}\label{eq:Sren}
  F_S(\mu) = Z_S^{-1} F_S^{\rm bare} \,. 
\end{align}
From the divergent terms in $F_{H}F_{JR}$ we may read off
\begin{align}
  Z_S = 1 + {\bar{\alpha}_0  \over 4\pi}{2\over \epsilon}\bigg( -\log{Q^2\over m^2} + 1 \bigg)
  + \left({\bar{\alpha}_0 \over 4\pi}\right)^2
    {2\over \epsilon^2}\bigg( -\log{Q^2\over m^2} + 1 \bigg)^2
      + \order(\alpha^3) \,,
\end{align}
with $\bar{\alpha}_0 = \alpha_0(\mu)$ the $\overline{\rm MS}$ coupling with $n_f=0$ (in $d=4$ dimensions, $\bar{\alpha}_0$ reduces
to the onshell $\alpha$).
The product of renormalized jet and remainder functions is given by
\begin{align}\label{eq:Rmu}
  F_{JR}(\mu)
  &= Z_H^{-1} Z_S F_{JR}^{\rm bare}
  \nl
  &= 1 + {\bar{\alpha}_1(\mu)\over 4\pi}\left( \log^2{m^2\over \mu^2} - \log{m^2\over\mu^2}
  + 4 + {\pi^2\over 6} \right)
  \nl
  &\quad
  + \left({\bar{\alpha}_1(\mu)\over 4\pi}\right)^2\bigg[
    \log{Q^2\over m^2}\bigg( -\frac43\log^2{m^2\over \mu^2} - {40\over 9}\log{m^2\over\mu^2}
    - {112\over 27}
    \bigg)
    + \frac12 \log^4{m^2\over\mu^2}
     \nl
    &\qquad 
    -\frac59 \log^3{m^2\over\mu^2}
   +\log^2{m^2\over\mu^2}\left( {53\over 18} + {\pi^2\over 6}\right)
    + \log{m^2\over\mu^2}\left( {251\over 54} + {49\pi^2\over 18} - 24\zeta_3 \right)
    \nl
    &\qquad
    -8\pi^2\log{2} + {76\pi^2\over 27}
    - {163\pi^4\over 360} - {58\zeta_3\over 9} + {39949\over 648}
    \bigg] + \order(\alpha^3) \,. 
\end{align}
The remaining (bare, unrenormalized) soft function for virtual corrections with nonvanishing $\lambda$ is 
\begin{align}\label{eq:S}
  \sqrt{ S(\Delta E=0)^{\rm bare} } \equiv F_S^{\rm bare} &= { F_1(q^2,m^2,\lambda^2) \over F_1(q^2,m^2,\lambda^2=0) } \,, 
\end{align}
where results for $F_1(q^2,m^2,\lambda^2)$ through two-loop order are given in Refs.~\cite{Kniehl:1989kz,Mastrolia:2003yz}. 
The renormalized soft function is given by Eq.~(\ref{eq:Sren}), or equivalently, 
\begin{align}\label{eq:Smu}
 F_S(\mu) &= { F_1(q^2,m^2,\lambda^2) \over F_H(\mu)F_{JR}(\mu) }
  \nl
  &= 
  1 + {\bar{\alpha}_0 \over 4\pi} \bigg[ 2\log{\lambda^2\over\mu^2}
    \bigg( \log{Q^2\over m^2} - 1 \bigg) \bigg]
  +  \left( {\bar{\alpha}_0 \over 4\pi} \right)^2 \bigg[ 2\log^2{\lambda^2\over\mu^2} 
    \bigg( \log{Q^2\over m^2} - 1 \bigg)^2 \bigg]
  + \order(\alpha^3) \,.
\end{align}

\subsection{Factorization of jet and remainder function}

Inspection of the explicit matching results in
Eqs.~(\ref{eq:Hmu}), (\ref{eq:Rmu}), and (\ref{eq:Smu})
reveals a pattern of large logarithms.
$H(\mu)$ is free of large logarithms provided $\mu \sim Q$.
$S(\mu)$ contains large logarithms irrespective of the choice of $\mu$,
but in an exponentiated form. 
The product $(JR)(\mu)$ is free of large logarithms through one loop order provided $\mu \sim m$, 
but contains large logarithms at two-loop order regardless of the choice of $\mu$ (except precisely $\mu=m$).

Note that the combinations $H$, $JR$ and $S$ are given by the simple momentum regions
analysis encoded by the form factor combinations in
Eqs.~(\ref{eq:hard}), (\ref{eq:R}) and (\ref{eq:S}), respectively.
A further factorization of the $JR$ function is obtained by considering an
intermediate effective theory in which the electron is dynamical inside closed
loops, but where the valence electron is treated as a heavy particle field.
The $R$ function is then given by matching the soft operator defined
in a theory with a dynamical fermion of mass $m$, to the soft operator defined
in a theory without dynamical fermion.  We find
\begin{align}
  \sqrt{R^{\rm bare}} &= F_R^{\rm bare} =
  1 + \left(\alpha_{\rm bare} m^{-2\epsilon} \over 4\pi \right)^{2}
  \bigg[ \log{Q^2\over m^2}\bigg( -{4\over 3\epsilon^2} + {20\over 9\epsilon} - {112\over 27}
    - {2\pi^2\over 9} \bigg)
    + {2\over\epsilon^2} - {8\over 3\epsilon} + {\pi^2\over 3}
    \nl
    &\quad
    + {52\over 9} \bigg]
  + \order(\alpha^3) \,,
\end{align}
where the result includes the two-loop vertex correction with closed fermion
loop~\cite{Becher:2007cu}, as well as a contribution from wavefunction renormalization
in the massive fermion theory~\cite{Broadhurst:1994se}.%
\footnote{
  This operator definition of $F_R$ differs from the quantity $\delta S$ in Ref.~\cite{Becher:2007cu}
  by inclusion of
  onshell renormalization factors.   The jet function $F_J$ in Eq.~(\ref{eq:FJren}) correspondingly differs from
  the quantity $Z_J$ in Ref.~\cite{Becher:2007cu}. 
}
After renormalization,
\begin{align}
  F_{R}(\mu) &= Z_{S} Z_{S,n_f=1}^{-1} F_R^{\rm bare}
  \nl
  &= 1 + \left( \bar{\alpha}_1(\mu) \over 4\pi \right)^2
  \bigg[ \log{Q^2\over m^2}\bigg( -\frac43 \log^2{m^2\over \mu^2}
    - \frac{40}{9}\log{m^2\over \mu^2} - {112\over 27} \bigg)
    + \frac83 \log^2{m^2\over \mu^2}
    \nl
    &\quad
    + \frac{16}{3}\log{m^2\over \mu^2}
    + {\pi^2\over 9} + {52\over 9}
    \bigg] + \order(\alpha^3) \,.
\end{align}
Having factored out $R(\mu)$, the remaining $J(\mu)$ is given by
\begin{align}\label{eq:FJren}
  \sqrt{J(\mu)} &= F_J(\mu)
  \nl &=
  1 +  {\bar{\alpha}_1(\mu)\over 4\pi}\left( \log^2{m^2\over \mu^2} - \log{m^2\over\mu^2}
  + 4 + {\pi^2\over 6} \right)
  + \left({\bar{\alpha}_1(\mu)\over 4\pi}\right)^2\bigg[
      \frac12 \log^4{m^2\over\mu^2}
    -\frac59 \log^3{m^2\over\mu^2}
  \nl
  &\quad
    +\log^2{m^2\over\mu^2}\left( {5\over 18} + {\pi^2\over 6}\right)
    + \log{m^2\over\mu^2}\left( -{37\over 54} + {49\pi^2\over 18} - 24\zeta_3 \right)
    -8\pi^2\log{2} + {73\pi^2\over 27}
    - {163\pi^4\over 360}
        \nl
    &\qquad
- {58\zeta_3\over 9} + {36205\over 648}
    \bigg]
  + \order(\alpha^3)  \,. 
\end{align}
Although the impact of $R(\mu)$ is numerically small, it is interesting
from a formal perspective 
to understand the all orders structure of large logarithms appearing in this function.
The operator definition identifying $R(\mu)$ as a ratio of Wilson loop matrix elements
in $n_f=1$ and $n_f=0$ can be used to show that $\log R(\mu)$ contains only a
single power of the large logarithm, $\log(Q^2/m^2)$, to all orders in perturbation
theory~\cite{Korchemsky:1987wg}.%
\footnote{
  In particular, $d \log R(\mu) / d\log\mu$ is given by the difference of cusp anomalous
  dimensions with $n_f=1$ and $n_f=0$, cf. Eqs.~(\ref{eq:hanom}),(\ref{eq:Gcusp}), and (\ref{eq:Gi}).
}
This ensures that high powers of large logarithms do not upset the power counting of the
resummed perturbative expansion.  
Such large logarithms have been studied in a variety of frameworks 
for applications involving massless fermions~\cite{Chiu:2007dg,Becher:2010tm,Chiu:2012ir}.%
\footnote{Reference~\cite{Chiu:2007dg} considers the massive fermion case through one loop order.}

\subsection{Soft-collinear factorization for real radiation} 

Factorization of the soft function in Eq.~(\ref{eq:facfull}) from the remaining process is
nontrivial.  It can be shown [cf. Eq.~(\ref{eq:Clead})] that multiple low-energy regions contribute to the
physical matrix element.  This complicates a simple eikonal decoupling argument like Eq.~(\ref{eq:decouple})
that applies in the heavy-particle case. 
Through two-loop order, factorization
is equivalent to the vanishing of additional contributions on the
right hand side of Eq.~(\ref{eq:deltahat}).  Direct evaluation of such
contributions is performed in the full theory in Appendix~\ref{sec:2loopfull}, 
and in the effective theory in 
Appendix~\ref{sec:2loopeff}.

\subsection{Two-loop soft function} 

Having derived the functions $H(\mu)$, $J(\mu)$ and $R(\mu)$, and having 
demonstrated soft-collinear factorization for real radiation, let
us specify the remaining soft function through two-loop order.
The complete soft function including real radiation,
$S(\Delta E)$ in Eq.~(\ref{eq:facfull}),
is obtained from Feynman diagrams with only soft photons, cf. Figs.~\ref{fig:scet} and \ref{fig:scet1}.
Our definition ensures that this function  
is identical to the soft function appearing in Eq.~(\ref{eq:fac_heavy}), extended to general $v\cdot v^\prime\gg 1$.%
\footnote{
  Note that with this definition, closed electron loop corrections are defined to be contained
  in $R$. 
  }
Using the explicit results (\ref{eq:Smu}) and (\ref{eq:double}), and the
soft contribution to Eq.~(\ref{eq:deltahat}), the complete corrections at one and two-loop
order are%
\footnote{
  The term $16\pi^2(L-1)^2 /3$ in $S^{(2)}$ has been noted in Ref.~\cite{Arbuzov:2015vba}. 
 }
\begin{align}\label{eq:fullsoft}
  S^{(1)} &= -4\left( \log{\mu^2\over m^2} + \log{E^2\over (\Delta E)^2}  \right)(L-1) + 2L^2
  + 4 {\rm Li}_2\left(  \cos^2{\theta\over 2} \right) - {4\pi^2\over 3} \,,
  \nl
  S^{(2)} &= \frac{1}{2!} [S^{(1)}]^2 - {16\pi^2 \over 3} (L-1)^2  \,.
\end{align}

\subsection{Effective theory: resummation}

After renormalization in the $\overline{\rm MS}$ scheme at scale $\mu$, 
the hard function is free of large logarithms provided that the matching
scale satisfies $\mu_H \sim Q$. 
Evolution to low scales $\mu_L \sim m$
is governed by (cf. Appendix~\ref{sec:RG})
\begin{align}\label{eq:Hanom}
  {d \log H \over d\log \mu}
  = 2\left[ \gamma_{\rm cusp}(\alpha) \log{Q^2\over \mu^2}  + \gamma(\alpha) \right] \,. 
\end{align}
The cusp anomalous dimension for massless QED ($n_f=1$) reads
\begin{align}\label{eq:gamcusp}
  \gamma_{\rm cusp} &= \sum_{n=0}^\infty \left(\bar{\alpha}\over 4\pi\right)^{n+1}
  \gamma^{\rm cusp}_n \,,
\qquad 
  \gamma^{\rm cusp}_0 = 4 
  \,, \quad
  \gamma^{\rm cusp}_1 = -{80\over 9} 
  \,.
\end{align}
The regular anomalous dimension $\gamma$ 
may be similarly expanded,
\begin{align}
  \gamma &= \sum_{n=0}^\infty \left(\bar{\alpha}\over 4\pi\right)^{n+1}
  \gamma_n \,, 
  \qquad 
  \gamma_0 = -6 \,.
\end{align}
Using these expansions, 
the solution of Eq.~(\ref{eq:Hanom}) to any order is straightforward.
Expressed in terms of the running coupling,
\begin{align}\label{eq:Hresum}
  &\log\left( {H(\mu_L)\over H(\mu_H)} \right)
  =
  -{\gamma_0\over \beta_0}\bigg\{  \log{r} + \dots \bigg\}
  -{\gamma^{\rm cusp}_0\over \beta_0} \bigg\{ \log{Q^2\over \mu_H^2}
  \log{r}
  +{1\over\beta_0}\bigg[ {4\pi \over\alpha(\mu_H)}\left( {1\over r} - 1+ \log{r} \right)
    \nl
    &\quad
    + \left( {\gamma^{\rm cusp}_1\over \gamma^{\rm cusp}_0} - {\beta_1\over\beta_0} \right)( -\log{r} + r - 1 )
  -{\beta_1\over 2\beta_0} \log^2{r}  \bigg]
  + \dots \bigg\}
   \,,
\end{align}
where $r = \alpha(\mu_L)/\alpha(\mu_H)$, and the first and second curly braces correspond
to the terms $\gamma(\alpha)$ and $\gamma_{\rm cusp}(\alpha)$ in Eq.~(\ref{eq:Hanom}), respectively. 

We are interested in applications involving large logarithms such that
$\alpha \log^2(\mu_H^2/\mu_L^2) \sim 1$.  In this power counting, terms involving
$\gamma_0$ scale as $\alpha^{1/2}$, and neglected terms involving $\gamma(\alpha)$
scale as $\alpha^{3/2}$.    The leading terms involving the cusp anomalous dimension
scale as $\alpha^0$, terms involving $\gamma^{\rm cusp}_1$ and $\beta_1$ scale as $\alpha^1$, and
the remaining neglected terms scale as $\alpha^2$.  
When combined with one-loop matching computations, the terms retained in Eq.~(\ref{eq:Hresum})
are thus sufficient to ensure accuracy through order $\alpha^1$, accounting for logarithmic enhancements.   
The result (\ref{eq:Hresum}) may be readily expressed in terms of the onshell coupling. 
Retaining terms through $\order(\alpha)$ in the above counting, 
\begin{align}\label{eq:Hresumexpand}
  &\log\left( {H(\mu_L)\over H(\mu_H)} \right)
  ={\alpha \over 4\pi}\bigg[  -2\log^2{\mu_H^2\over \mu_L^2} -4\log{\mu_H^2\over\mu_L^2}\log{Q^2\over\mu_H^2}
+ 6\log{\mu_H^2\over\mu_L^2} 
    \bigg]  
  \nl
  &\qquad
  + \left(\alpha\over 4\pi\right)^2
  \bigg[ -\frac89 \log^3{\mu_H^2 \over \mu_L^2}
    -\frac83 \log^2{\mu_H^2\over\mu_L^2} \left( \log{Q^2\over \mu_H^2} - \log{m^2\over \mu_L^2} \right)
    + {76\over 9} \log^2{\mu_H^2\over \mu_L^2} + \dots 
    \bigg]
  \nl
  &\qquad
  + \left(\alpha\over 4\pi\right)^3
  \bigg[ {176\over 27} \log^4{\mu_H^2\over\mu_L^2} + \dots \bigg]  + \dots \,. 
\end{align}

\begin{figure}[tb]
  \begin{center}
    \includegraphics[width=0.6\textwidth]{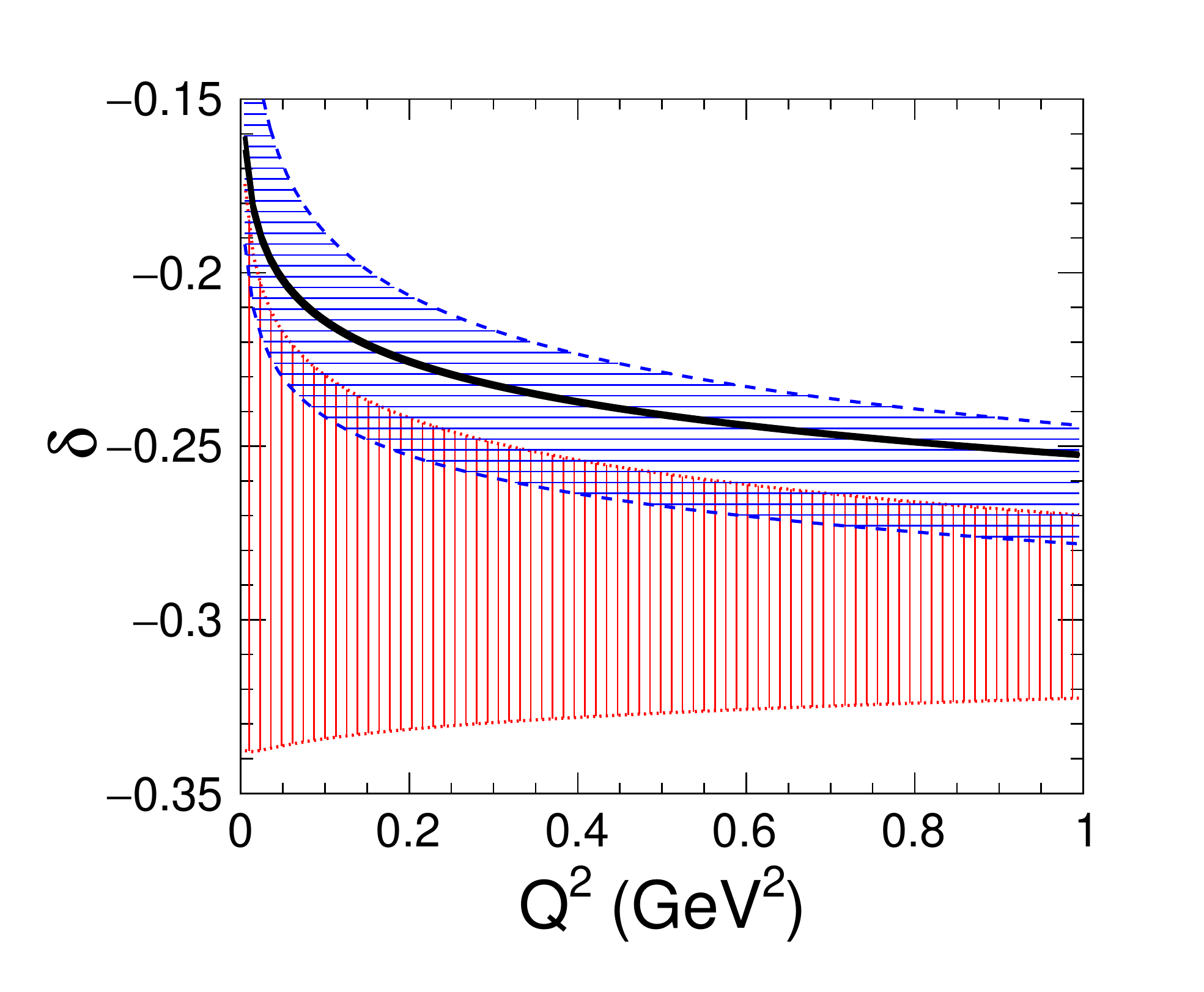}
    \vspace{-8mm}
    \caption{      \label{fig:error_static}
      Radiative correction factor $\delta$ in resummed perturbation theory
      for the static source limit of electron-proton scattering,
      with $E=1\,{\rm GeV}$, $\Delta E= 5\,{\rm MeV}$.
       The bands represent the impact of varying
      ${\rm min}(Q^2,E^2)/2 < \mu_H^2 < 2\,{\rm max}(Q^2,E^2)$
       and ${\rm min}(m^2,\Delta E^2)/2 < \mu_L^2 < 2\,{\rm max}(m^2,\Delta E^2)$, 
       using 
       leading log resummation (blue, horizontal stripes)
       next-to-leading log resummation (red, vertical stripes) 
       and complete next-to-leading order resummation (black, solid band).  
    }
  \end{center}
\end{figure}

With the result (\ref{eq:Hresumexpand}), we have control over large logarithms
and a complete solution through true order $\alpha$ (i.e., all neglected terms
are parametrically small compared to order $\alpha$, accounting for logarithmic enhancements).  
Setting $\mu_L \sim m$, inspection of $S(\mu_L)$ shows that the non-exponentiating term in $S^{(2)}$ is of
order $\alpha^2 L^2 \sim \alpha^1$.  $J(\mu_L)$ contains no large logarithms and may be truncated at one-loop order.  
$R(\mu_L)$ is nontrivial only at order $\alpha^{3/2}$, and may be neglected.
Similarly, setting $\mu_H \sim M$, the matching coefficient $H(\mu_H)$ is free of
large logarithms and may be truncated at one-loop order.    
Figure~\ref{fig:error_static} compares successive inclusion of
terms at order $\alpha^0$, $\alpha^{\frac12}$ and $\alpha^1$ in resummed perturbation theory.
The figure demonstrates the necessity to control both leading and subleading
logarithms in the perturbative expansion.

\subsection{Nuclear recoil and structure corrections}

The preceding discussion gives a complete solution including subleading log resummation
for the idealized problem of scattering from a static source. 
Let us include the effects of nuclear recoil and structure. 
The ``Born'' cross section (denoted with subscript $0$) is~\cite{Lee:2015jqa} 
\be
(d\sigma)_0 = (d\sigma)_{\rm Mott} { \epsilon G_E^2 + \tau G_M^2 \over \epsilon (1+\tau)} \,,
\ee
where the Mott cross section is now
$(d\sigma/d\Omega)_{\rm Mott} = \alpha^2 \cos^2(\theta/2)/[4 \eta E^2 \sin^4(\theta/2)]$,
with
\be
\eta = E/E^\prime\,,\quad \tau = {Q^2\over 4M^2} \,, \quad
\epsilon^{-1} = 1 + 2(1+\tau)\tan^2{\theta\over 2} \,.
\ee
To begin, we work to first order in nuclear charge, i.e., neglect
radiative corrections involving the proton. 
The experimentally measured cross section is
\be
d\sigma = { (d\sigma)_{0}  \over [1-\hat{\Pi}(q^2)]^2 }
\left( 1 + \delta_{e} + \delta_{e\gamma} + \delta_{e\gamma\gamma} + \dots
\right) \,. 
\ee
The virtual corrections as a function of $q^2$ are identical to
the static case,
\be\label{eq:delta0_rec}
\delta_e = |F_q(q^2,m^2,\lambda^2)|^2 - 1 \,.
\ee
First order real corrections are now~\cite{Maximon:2000hm}
\begin{align}\label{eq:delta1_rec}
  \delta^{(1)}_{e\gamma}
  &= 4\left( \log{ (\eta \Delta E)^2 \over EE^\prime} -\log{\lambda^2\over m^2} \right)(L-1)
  +2 L^2
  -2 \log^2\eta + 4{\rm Li}_2\left(\cos^2{\theta\over 2}\right) - {4\pi^2\over 3} \,. 
\end{align}
In terms of this result, second order real corrections are
\be\label{eq:delta2_rec}
\delta^{(2)}_{e\gamma\gamma} = {1\over 2!} \left[ \delta^{(1)}_{e\gamma} \right]^2
- {16 \pi^2\over 3} (L-1)^2 \,. 
\ee
Assuming soft-collinear factorization, the mixed real-virtual contribution
at second order is given in terms of the result (\ref{eq:delta1_rec}) by
\begin{align}\label{eq:delta3_rec}
  \delta_{e\gamma}^{(2)} &=
 \delta^{(1)}_e  \delta_{e\gamma}^{(1)} \,.
\end{align}
The results (\ref{eq:delta0_rec}), (\ref{eq:delta1_rec}), (\ref{eq:delta2_rec})
and (\ref{eq:delta3_rec})
imply that Eq.(\ref{eq:tot}) remains valid when recoil effects are included.  

\subsection{Two photon exchange}

The complete result at first order in nuclear charge is simplified by
the factorization theorem which implies that recoil effects
are confined to soft function contributions involving real emission.
Beyond first order in the nuclear charge, radiative corrections introduce
new operators at the hard scale, and sensitivity to nuclear structure beyond form factors.
Let us briefly discuss the inclusion of such corrections in the formalism.

The factorization formula including second (and higher) order corrections
in nuclear charge takes the same form as Eq.~(\ref{eq:facfull}).   The function $J(\mu)$
is unchanged.   The function $R(\mu)$ may be taken as unity at the relevant order
[recall $R \sim \alpha^2 L =\order(\alpha^{3/2})$ in our counting $\alpha L^2 =\order(1)$] .
Let us focus on the hard and soft functions.  In particular, let us consider the
extraction of proton structure information from scattering data.  Our
goal is to isolate $H(\mu=M)$, which is built from conventionally defined Born
form factors, as in Eq.~(\ref{eq:born}), and analogous hard coefficient functions
arising from two-photon exchange.
In the absence of sufficient data~\cite{Rimal:2016toz} to simultaneously extract the Born form factors
and the two-photon exchange contributions to $H(\mu=M)$, hadronic models are employed
for the latter~\cite{Blunden:2003sp,Carlson:2007sp}.

The soft function (as well as the remainder function $R$ and jet function $J$) is universal to all
of the underlying amplitudes.   In place of the static-source limit of Eq.~(\ref{eq:FS}),
we have now
\begin{align}\label{eq:softTPE}
\sqrt{S(\mu,\Delta E=0)} &= 
  Z_h^{(e)} Z_h^{(p)} \left|
  \parbox{25mm}{
      \begin{fmfgraph*}(70,40)
        \fmfset{arrow_len}{3mm}\fmfset{arrow_ang}{12}
        \fmfleftn{l}{3}
        \fmfrightn{r}{3}
        \fmf{phantom}{l3,v,r3}
        \fmf{phantom}{l1,v,r1}
        \fmffreeze
        \fmf{double}{l3,v,r3}
        \fmf{double}{l1,v,r1}
        \fmfv{d.sh=circle,d.f=empty,d.si=.1w,l=$\times$,label.dist=0}{v}
      \end{fmfgraph*}
    }
    +
  \parbox{25mm}{
      \begin{fmfgraph*}(70,40)
        \fmfset{arrow_len}{3mm}\fmfset{arrow_ang}{12}
        \fmfleftn{l}{3}
        \fmfrightn{r}{3}
        \fmf{phantom}{l3,x,v,y,r3}
        \fmf{phantom}{l1,a,v,b,r1}
        \fmffreeze
        \fmf{photon,left}{x,y}
        \fmf{double}{l3,v,r3}
        \fmf{double}{l1,v,r1}
           \fmfv{d.sh=circle,d.f=empty,d.si=.1w,l=$\times$,label.dist=0}{v}
      \end{fmfgraph*}
  }
    +
  \parbox{25mm}{
      \begin{fmfgraph*}(70,40)
        \fmfset{arrow_len}{3mm}\fmfset{arrow_ang}{12}
        \fmfleftn{l}{3}
        \fmfrightn{r}{3}
        \fmf{phantom}{l3,x,v,y,r3}
        \fmf{phantom}{l1,a,v,b,r1}
        \fmffreeze
        \fmf{photon,left}{b,a}
        \fmf{double}{l3,v,r3}
        \fmf{double}{l1,v,r1}
           \fmfv{d.sh=circle,d.f=empty,d.si=.1w,l=$\times$,label.dist=0}{v}
      \end{fmfgraph*}
  }
  +
  \parbox{25mm}{
      \begin{fmfgraph*}(70,40)
        \fmfset{arrow_len}{3mm}\fmfset{arrow_ang}{12}
        \fmfleftn{l}{3}
        \fmfrightn{r}{3}
        \fmf{phantom}{l3,x,v,y,r3}
        \fmf{phantom}{l1,a,v,b,r1}
        \fmffreeze
        \fmf{photon,left}{a,x}
        \fmf{double}{l3,v,r3}
        \fmf{double}{l1,v,r1}
           \fmfv{d.sh=circle,d.f=empty,d.si=.1w,l=$\times$,label.dist=0}{v}
      \end{fmfgraph*}
  }
  \right.
  \nonumber
  \\[2mm]
&\quad 
\left.
  +
  \parbox{25mm}{
      \begin{fmfgraph*}(70,40)
        \fmfset{arrow_len}{3mm}\fmfset{arrow_ang}{12}
        \fmfleftn{l}{3}
        \fmfrightn{r}{3}
        \fmf{phantom}{l3,x,v,y,r3}
        \fmf{phantom}{l1,a,v,b,r1}
        \fmffreeze
        \fmf{photon,left}{y,b}
        \fmf{double}{l3,v,r3}
        \fmf{double}{l1,v,r1}
           \fmfv{d.sh=circle,d.f=empty,d.si=.1w,l=$\times$,label.dist=0}{v}
      \end{fmfgraph*}
  }
  +
  \parbox{25mm}{
      \begin{fmfgraph*}(70,40)
        \fmfset{arrow_len}{3mm}\fmfset{arrow_ang}{12}
        \fmfleftn{l}{3}
        \fmfrightn{r}{3}
        \fmf{phantom}{l3,x,v,y,r3}
        \fmf{phantom}{l1,a,v,b,r1}
        \fmffreeze
        \fmf{photon,left}{x,b}
        \fmf{double}{l3,v,r3}
        \fmf{double}{l1,v,r1}
           \fmfv{d.sh=circle,d.f=empty,d.si=.1w,l=$\times$,label.dist=0}{v}
      \end{fmfgraph*}
  }
  +
  \parbox{25mm}{
      \begin{fmfgraph*}(70,40)
        \fmfset{arrow_len}{3mm}\fmfset{arrow_ang}{12}
        \fmfleftn{l}{3}
        \fmfrightn{r}{3}
        \fmf{phantom}{l3,x,v,y,r3}
        \fmf{phantom}{l1,a,v,b,r1}
        \fmffreeze
        \fmf{photon,left}{a,y}
        \fmf{double}{l3,v,r3}
        \fmf{double}{l1,v,r1}
           \fmfv{d.sh=circle,d.f=empty,d.si=.1w,l=$\times$,label.dist=0}{v}
      \end{fmfgraph*}
  }
  \right|
  \nl
  &= 1 - {\alpha\over 2\pi} {\rm Re}\bigg\{
  \big[ u\cdot u^\prime f(u\cdot u^\prime) - 1 \big]
  + Z^2 \big[ v\cdot v^\prime f(v\cdot v^\prime) - 1 \big]
  \nl
  &\quad 
  + Z \big[ u\cdot v f(-u\cdot v - i0) + u^\prime \cdot v^\prime f(-u^\prime \cdot v^\prime-i0)
    + u \cdot v^\prime f(u \cdot v^\prime)
    \nl
    &\quad
    + u^\prime \cdot v f(u^\prime\cdot v)
    \big]\bigg\}
  \log{\mu^2\over \lambda^2} 
  \,,  
\end{align}
where $u^\mu$, $u^{\prime \mu}$ are timelike vectors proportional to initial and
final electron momentum, and $v^\mu$, $v^{\prime\mu}$ similarly correspond to the
momenta of the initial and final state proton.
The function $f(w)$ was introduced for $w\ge 1$ in Eq.~(\ref{eq:rw}), and the explicit
evaluation of the Feynman integrals yields
\be
f(-w-i0) = -f(w) + {i\pi \over\sqrt{w^2-1}} \,. 
\ee
The kinematic constraints,
\be
v^\prime \cdot u = v\cdot u^\prime \,, \quad
v^\prime \cdot u^\prime = v\cdot u \,,
\ee
may be used to reduce the number of terms appearing in Eq.~(\ref{eq:softTPE}).

\begin{figure}[tb]
  \begin{center}
    \includegraphics[width=0.6\textwidth]{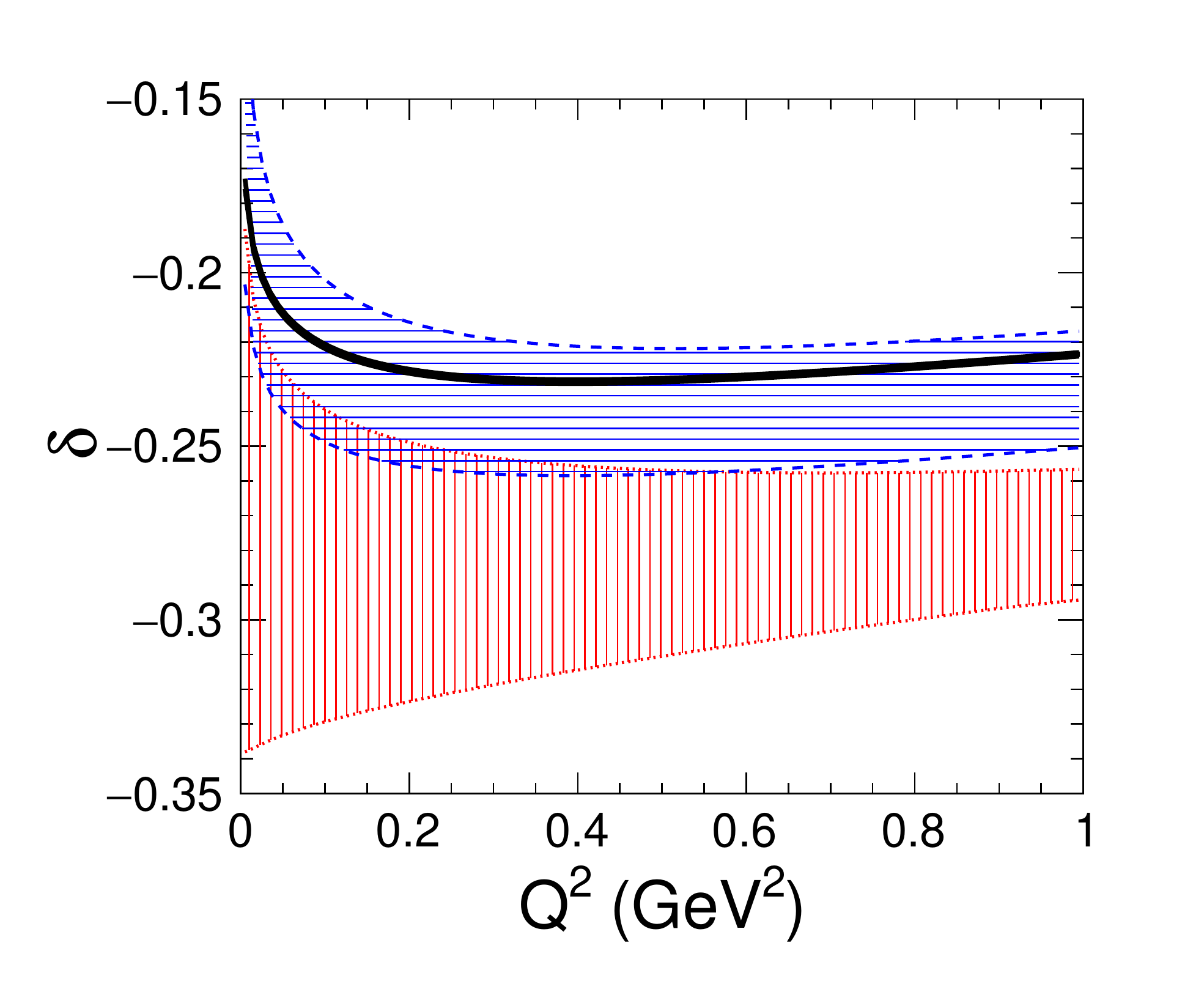}
    \caption{  \label{fig:error_full}
      Same as Fig.~\ref{fig:error_static}, but including recoil
      and nuclear charge corrections (i.e., two photon exchange and
      proton vertex corrections).  
    }
  \end{center}
\end{figure}

In order to extract the hard function at scale $\mu=M$, we write the process as
\be
d\sigma \propto H(M) \times  {H(\mu) \over H(M) } \times (JRS)(\mu) \,,
\ee
evaluating $JRS$ at the soft scale, and thus requiring the ratio $H(\mu)/H(M)$,
with control over large logarithms in perturbation theory.
The renormalization of the hard function is now governed by (cf. Appendix~\ref{sec:RG})
\begin{align}\label{eq:Hfullresum}
  {d\log H\over d\log\mu} &= 2\bigg[ \gamma_{\rm cusp}(\bar{\alpha}) \log{Q^2\over\mu^2}
    + \gamma_{\rm cusp}(v\cdot v^\prime, \bar{\alpha})
    + 2 \gamma_{\rm cusp}(\bar{\alpha}) \log{v\cdot p^\prime \over -v\cdot p-i0}
    + \gamma(\bar{\alpha}) 
    \bigg] \,.
\end{align}
The cusp function $\gamma_{\rm cusp}(\bar{\alpha})$ has been 
introduced above in Eq.~(\ref{eq:Hanom}), $\gamma_{\rm cusp}(w,\bar{\alpha})$ is
given in Eq.~(\ref{eq:massivecusp}),  and 
the regular anomalous dimension $\gamma(\bar{\alpha})$ is
\begin{align}
  \gamma &= \sum_{n=0}^\infty \left(\bar{\alpha}\over 4\pi\right)^{n+1}
  \gamma_n \,, 
  \qquad 
  \gamma_0 = -10 \,.
\end{align}
The solution to Eq.~(\ref{eq:Hfullresum}), analogous to Eq.~(\ref{eq:Hresum}), is
\begin{align}
  &\log {H(\mu_L)\over H(\mu_H)}
  =
  -{1\over \beta_0} \bigg[ \gamma_0 +  \left( \log{Q^2\over \mu_H^2} +   wf(w) +  2\log{E^\prime \over -E-i0} \right) \gamma^{\rm cusp}_0
        \bigg]\log{r}
  \nl&\quad
  -{\gamma^{\rm cusp}_0\over \beta_0^2} \bigg\{
  {4\pi \over\alpha(\mu_h)}\left( {1\over r} - 1+ \log{r} \right)
    + \left( {\gamma^{\rm cusp}_1\over \gamma^{\rm cusp}_0} - {\beta_1\over\beta_0} \right)( -\log{r} + r - 1 )
  -{\beta_1\over 2\beta_0} \log^2{r} 
  + \dots \bigg\}
  \,.
\end{align}
Expressed in terms of onshell coupling, 
\begin{align}\label{eq:completeresum}
  \log {H(\mu_L)\over H(\mu_H)}
  &=\bigg\{ {\alpha \over 4\pi}\bigg[  -2\log^2{\mu_H^2\over \mu_L^2} -4\log{\mu_H^2\over\mu_L^2}\log{Q^2\over\mu_H^2}
    \bigg]  
  \nl
  &\quad
  + \left(\alpha\over 4\pi\right)^2
  \bigg[ -\frac89 \log^3{\mu_H^2 \over \mu_L^2}
    -\frac83 \log^2{\mu_H^2\over\mu_L^2} \left( \log{Q^2\over \mu_H^2} - \log{m^2\over \mu_L^2} \right)
    + {40\over 9} \log^2{\mu_H^2\over \mu_L^2} + \dots 
    \bigg]
  \nl
  &\quad
  + \left(\alpha\over 4\pi\right)^3
  \bigg[ {176\over 27} \log^4{\mu_H^2\over\mu_L^2} + \dots \bigg]  + \dots \bigg\}
  +
  \bigg[ -10 + 4 w f(w) + 8 \log{E^\prime\over -E-i0} \bigg]
  \nl
  &\quad
  \times 
  \bigg\{ {\alpha\over 4\pi}\bigg[ -\log{\mu_H^2\over \mu_L^2}  \bigg] + \left(\alpha\over 4\pi\right)^2\bigg[ -\frac23 \log^2{\mu_H^2\over\mu_L^2} + \dots \bigg]
  + \dots \bigg\} \,.  
\end{align}
where terms through $\alpha^1$ are retained, in the counting $\alpha \log^2(Q^2/m^2) \sim 1$.  
The impact of successive terms in the resummed perturbative expansion is displayed in Fig.~\ref{fig:error_full}. 

\section{Discussion \label{sec:discuss}} 

\begin{figure}[tb]
  \begin{center}
    \includegraphics[width=0.60\textwidth]{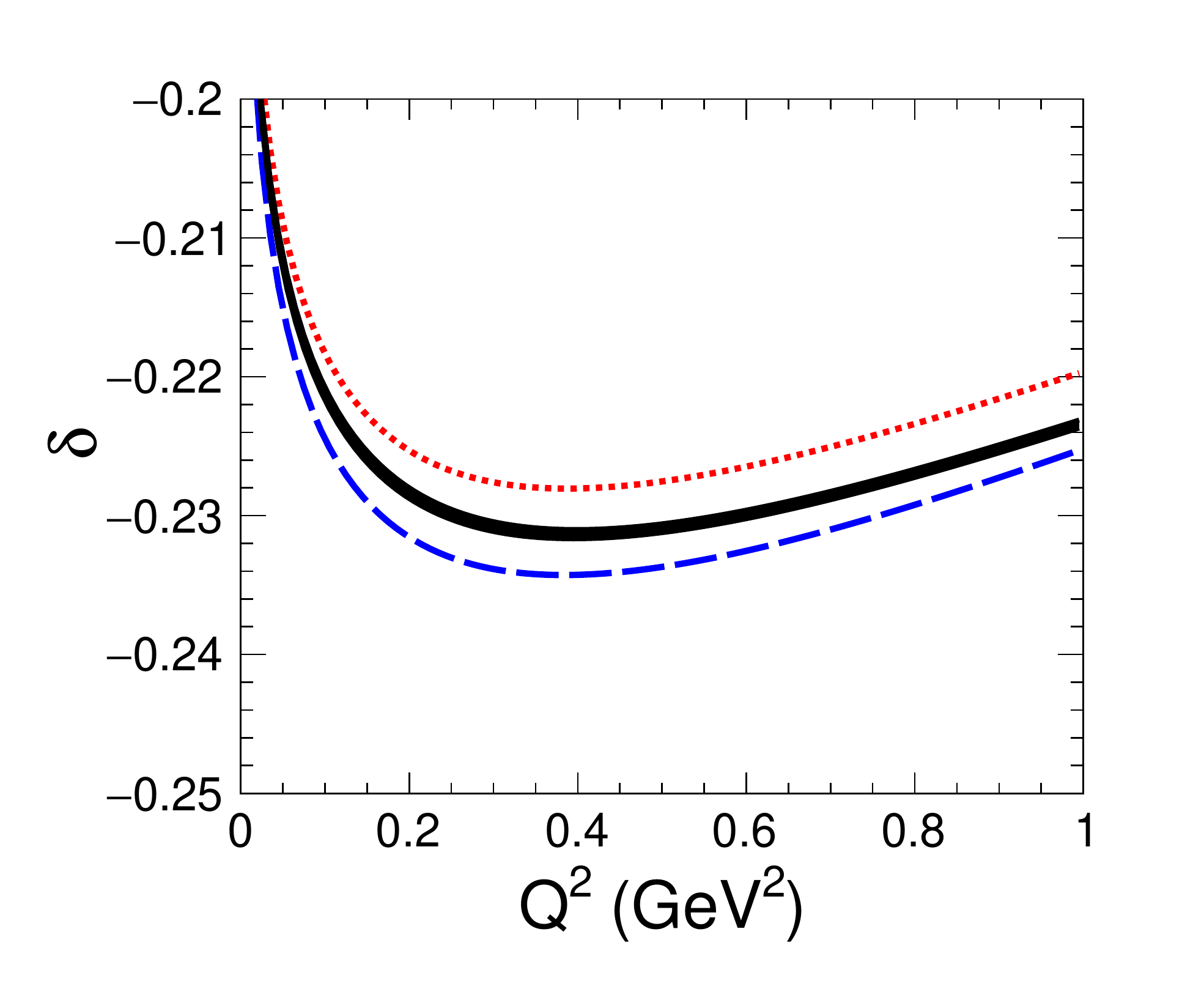}
    \caption{  \label{fig:resum_v_exp}
      Comparison of complete next to leading order resummed correction (solid black band) to
      naive exponentiations using different factorization scales for the two photon exchange
      correction: $\mu^2=M^2$ (dotted red line) and $\mu^2=Q^2$ (dashed blue line).
      See text for details. 
    }
  \end{center}
\end{figure}

The precision of electron-proton scattering experiments has reached a level
demanding systematic analysis of subleading radiative corrections at two loop
order and beyond.   We have presented the general framework that separates
physical scales in the scattering process, allowing a systematic merger
of fixed order perturbation theory with large log resummation. 

The quantum field theory analysis reveals implicit conventions and assumptions
that often differ between applications, such as between scattering and bound state
problems.   The definition of the proton charge and magnetic radii in the
presence of electromagnetic radiative corrections is naturally defined
in Eq.~(\ref{eq:born}).  A comparison to other definitions in the literature
is presented in Appendix~\ref{sec:convent}.
The separation of soft and hard scales in two photon exchange is similarly
ambiguous in standard treatments.  The common Maximon-Tjon convention~\cite{Maximon:2000hm} implicitly
takes momentum-dependent factorization scale $\mu^2=Q^2$ for two-photon exchange, in conflict
with the $Q^2$-independent choice $\mu^2=M^2$ that is closest to the implicit
convention for vertex corrections.

The exponentiation and cancellation of infrared singularities~\cite{Yennie:1961ad} in physical
processes has often been used to motivate a simple exponentiation of
first order corrections in order to resum logarithmically enhanced radiative
corrections at second- and higher-order in perturbation theory~\cite{Vanderhaeghen:2000ws,Bernauer:2013tpr}.
This procedure fails to capture subleading logarithms, beginning at order
$\alpha^2 L^3 = \order( \alpha^\frac12)$, in our counting $\alpha L^2 = \order(1)$, 
cf. Eq.~(\ref{eq:tot}).  These large logarithms are automatically generated
in the renormalization analysis that the effective theory makes possible.
The convergence of resummed perturbation theory is illustrated, for the complete problem
including proton structure and recoil, in Fig.~\ref{fig:error_full}.  
A comparison of the resummed prediction to the naive exponentiation ansatz
is displayed in Fig.~\ref{fig:resum_v_exp}.

Also shown in Fig.~\ref{fig:resum_v_exp} is the variation due to different scale
choices implicit in different two-photon exchange corrections.%
\footnote{
  For example, the so-called McKinley-Feshbach correction~\cite{McKinley:1948zz}
  represents the large-$M$ limit of the hard-coefficient contribution to two-photon exchange,
  and is independent of factorization scale $\mu$. 
  Using this correction~\cite{Bernauer:2013tpr} results in an irreducible factorization-scale
  uncertainty, uncanceled between matrix element and coefficient.
}
These ansatzes differ at the percent level in the considered kinematic range, and
fall well outside the error band represented by the complete next-to-leading
order resummed prediction.

Special attention has been paid to the effects of real emission beyond tree level.
Soft-photon factorization and exponentiation is readily proven~\cite{Yennie:1961ad} for the case
$\Delta E \ll m$.  In practical experiments, the opposite limit, $m\ll \Delta E$, obtains.
It is readily seen (cf. Appendix~\ref{sec:2loopfull}) that multiple low-energy momentum
regions appear, invalidating a simple factorization argument.   Nevertheless, an explicit
computation of the two-loop mixed real-virtual correction demonstrates factorization
for the simplest elastic scattering observable under consideration.
Extensions to
other observables, including the possibility of hard photon emission,
will be considered elsewhere. 

Discrepancies at the $0.5 - 1\%$ level exist between the complete resummed prediction
(\ref{eq:completeresum}), and phenomenological approximations employed in the analysis of
A1 collaboration electron-proton scattering data~\cite{Bernauer:2013tpr}, 
as illustrated in Fig.~\ref{fig:resum_v_exp}.  
It is interesting to consider the impact of these corrections on the proton radius puzzle. 
These discrepancies are in tension with the $0.2 - 0.5\%$ systematic errors
assumed in the extraction of proton electric and magnetic charge radii~\cite{Bernauer:2013tpr}, 
but will be partially absorbed by floating normalization parameters in fits
to independent datasets~\cite{Bernauer:2013tpr,Lee:2015jqa}.  A careful accounting of correlated
shape variations induced by radiative corrections 
must also be accounted for when fitting the inferred radiative tail for the
signal process together with background processes~\cite{Sick:2012zz,Bernauer:2013tpr}. 
The complete implementation of improved corrections in the analysis of
electron-proton scattering data, for charge radius and form factor extractions, is outside
the scope of this paper~\cite{newpaper}.
It is straightforward to include these improvements in event
generators~\cite{Vanderhaeghen:2000ws,Actis:2010gg,Jegerlehner:2011mw,Gramolin:2014pva}. 
It is interesting to perform a systematic analysis of power corrections in this framework, particularly
of relevance to very low $Q^2$ and/or high $\Delta E$~\cite{Akushevich:2015toa,Gramolin:2016hjt}.%
\footnote{First order power corrections in the static source limit are obtained from 
  the integrals in Sec.~\ref{sec:phase}.  These are small in the kinematics of the A1 collaboration
data~\cite{Bernauer:2013tpr}.}

Many other lepton-hadron processes are being probed at the percent and permille level, 
and are critical to next generation experiments probing fundamental physics
in and beyond the standard model.  Examples include neutrino-nucleus scattering for
neutrino oscillations~\cite{Day:2012gb}, $e^+e^- \to {\rm hadrons}$ for input to $(g-2)_\mu$~\cite{Lees:2015qna},
and parity violating scattering observables~\cite{Benesch:2014bas,Aleksejevs:2012sq,Aleksejevs:2015dba}. 
The effective field theory analysis may be readily applied to systematically
compute radiative corrections involving large logarithms in these and other
applications.

\vskip 0.2in
\noindent
 {\bf Acknowledgements.}
 The author thanks J. Arrington and G. Lee for collaboration on Ref.~\cite{Lee:2015jqa}
 which motivated the present work, and T.~Becher, G.~Paz and J.~Sapirstein for comments
 on the manuscript.  
 Research supported by a NIST Precision Measurement Grant and the U.S. Department of Energy,
 Office of Science, Office of High Energy Physics (DOE Grant No. DE-FG02-13ER41958).
 TRIUMF receives federal funding via a contribution agreement with the National Research Council of Canada.
 Research at Perimeter Institute is supported by the Government of Canada through the
 Department of Innovation, Science and Economic Development and by the Province of
 Ontario through the Ministry of Research and Innovation. 
\vskip 0.1in
\noindent

\appendix

\section{Renormalization constants \label{sec:RG} }

We collect here standard renormalization constants and conventions used in the paper.
Working in $d=4-2\epsilon$ dimensions, the bare QED coupling $e_{\rm bare}$ and fine structure constant $\alpha_{\rm bare}$
are defined and related to the $\overline{\rm MS}$ fine structure constant $\bar{\alpha}\equiv \alpha(\mu)$ by
\begin{align}
   {e_{\rm bare}^2\over 4\pi} (4\pi)^\epsilon e^{-\gamma_E \epsilon} = \alpha_{\rm bare} = \mu^{2 \epsilon}\bar{\alpha}
   \left[ 1 + \sum_{n=0}^\infty Z_n \left(\bar{\alpha}\over 4\pi\right)^{n+1} \right]\,,
   \quad
   Z_0 = {4\over 3\epsilon} \,. 
\end{align}
   The QED beta function is defined as
\begin{align}
  {d \bar{\alpha}\over d\log\mu} = -2\bar{\alpha} \sum_{n=0}^\infty \beta_n \left(\bar{\alpha}\over 4\pi\right)^{n+1} \,,
  \quad 
  \beta_0 = -\frac43 n_f \,, \quad
  \beta_1 = -4 n_f \,. 
\end{align}
The relation between onshell and $\overline{\rm MS}$ couplings with $n_f=1$ light flavors of mass $m$ is 
(in $d=4$)
\begin{align}
  \bar{\alpha} \equiv \alpha(\mu) = \alpha \left[ 1 + \sum_{n=0}^\infty z_n \left(\alpha\over 4\pi\right)^{n+1} \right]
  \,, \quad
  z_0 = \frac83 \log{\mu\over m} \,, \quad
  z_1 = \frac{64}{9} \log^2{\mu\over m} + 8\log{\mu\over m} + 15 \,.  
\end{align}
The onshell wavefunction renormalization constants for massive relativistic (QED) and nonrelativistic (NRQED)
fermions are
\be\label{eq:Zonshell}
Z_\Psi = 1 + {\bar{\alpha}\over 4\pi} \left( -{1\over \epsilon} + \log{m^2\over \mu^2} - 2\log{\lambda^2\over m^2} - 4 \right) \,,
\quad 
Z_h = 1 + {\bar{\alpha}\over 4\pi} \left( {2\over\epsilon} - 2\log{\lambda^2\over \mu^2} \right) \,. 
\ee

Consider the renormalization of Wilson coefficients for operators representing the soft and collinear
matrix elements for physical amplitudes specified by external momenta of a given collection
of massless and massive fermions.  Let the massless ($\psi$) and massive ($h$) fermions be labeled by lowercase
indices ${i}$, and uppercase indices ${I}$, respectively. 
In general,~\cite{Becher:2003kh,Becher:2009qa,Becher:2009kw,Beneke:2009rj} 
\begin{align}
  {d \log C \over d \log \mu} &= 
  \sum_{\{i,j\}} Q_i Q_j \gamma_{\rm cusp}(\bar{\alpha}) \log{\mu^2\over-s_{ij}}
  -
  \sum_{\{I,J\}} Q_I Q_J \gamma_{\rm cusp}\left({-s_{IJ}\over M_I M_J},\bar{\alpha} \right)
  \nl
  &\quad
  +
  \sum_{\{I,j\}} Q_I Q_j \gamma_{\rm cusp}(\bar{\alpha}) \log{ M_I \mu \over-s_{Ij}}
  + \sum_{i} \gamma^h(\bar{\alpha})
  + \sum_{I} \gamma^\psi(\bar{\alpha}) \,, 
\end{align}
where sums $\{i,j\}$ run over sets of distinct particle indices.
Here $Q_i$ denotes the electric charge (in units of the proton charge)
of the fermion, with all lines in a Feynman diagram viewed as ingoing
(so, e.g., $Q_i=-1$ for an incoming electron, $Q_i=+1$ for an outgoing electron).
Also, $s_{ij} = 2p_i\cdot p_j + i0$, where all momenta are viewed as incoming.   

Here the massless cusp function is 
\begin{align}\label{eq:cuspeq}
  \gamma_{\rm cusp}(\bar{\alpha}) &= \sum_{n=0}^\infty \left(\bar{\alpha}\over 4\pi\right)^{n+1}
  \gamma^{\rm cusp}_n \,,
\qquad 
  \gamma^{\rm cusp}_0 = 4 
  \,, \quad
  \gamma^{\rm cusp}_1 = -{80\over 9} n_f 
  \,.
\end{align}
The massive cusp function is
\begin{align}\label{eq:massivecusp}
  \gamma_{\rm cusp}(w,\bar{\alpha}) = \gamma_{\rm cusp}(\bar{\alpha}) w f(w) \,, 
\end{align}
with $f(w)$ as in Eq.~(\ref{eq:rw}) and $\gamma_{\rm cusp}(\bar{\alpha})$ as in Eq.~(\ref{eq:cuspeq}). 
The one-particle terms for massless fermions are
\begin{align}
\gamma^\psi &= \sum_{n=0}^\infty \left(\bar{\alpha}\over 4\pi\right)^{n+1}
  \gamma^\psi_n \,,
\qquad 
  \gamma^\psi_0 = -3  
  \,,
\end{align}
while for massive fermions 
\begin{align}
\gamma^h &= \sum_{n=0}^\infty \left(\bar{\alpha}\over 4\pi\right)^{n+1}
  \gamma^h_n \,,
\qquad 
  \gamma^h_0 = -2 \,, \quad \gamma^h_1 = {40\over 9}n_f \,. 
\end{align}
With these general results, we obtain the anomalous dimensions for hard functions
in Eqs.~(\ref{eq:hanom}), (\ref{eq:Hanom}) and (\ref{eq:Hfullresum}).
In particular, in Eq.~(\ref{eq:hanom}) we identify
$\Gamma_{\rm cusp}(w,\bar{\alpha})=\gamma_{\rm cusp}(w,\bar{\alpha}) + 2\gamma^h(\bar{\alpha})$.
In Eq.~(\ref{eq:Hanom}) we identify $\gamma=2\gamma^\psi$, and
in Eq.~(\ref{eq:Hfullresum}) we identify $\gamma=2\gamma^\psi + 2\gamma^h$.   

\section{Born conventions \label{sec:convent}}

A number of conflicting conventions exist in the electron-proton
scattering literature for defining infrared finite Born form factors.
These must all be of the form, 
\begin{align}
F_i(q^2)^{\rm Born} \equiv \tilde{F}_i(q^2)\bigg\{ 1
-
{Z^2 \alpha \over 2\pi}\bigg[ ( w f(w) -1 )\log{M^2\over \lambda^2}
  + \Delta K \bigg] + \order(\alpha^2)
\bigg\} \,,
\end{align}
as derived in the effective theory analysis.  
Here $\tilde{F}_i$ denotes the onshell form factor, and
$F_i^{\rm Born}(0)= \tilde{F}_i(0)$.
Several conventions are listed here for the finite term $\Delta K$.
The natural convention based on the factorization formulas discussed in this
paper is 
\begin{align}\label{eq:Kfac}
  \Delta K^{\rm fac.} &= 0 \,.
\end{align}
The convention adopted in Ref.~\cite{Lee:2015jqa} is essentially that
of Maximon and Tjon~\cite{Maximon:2000hm}, but neglecting an
additional model-dependent correction (referred to as $\delta^{(1)}_{\rm el}$ in
Ref.~\cite{Maximon:2000hm}), 
\begin{align}\label{eq:LAH}
  \Delta K^{\rm LAH} &= {w\over\sqrt{w^2-1}} \bigg[
    \log w_+\log[2(w+1)] - 2 {\rm Li}_2\left(\frac{-1}{w_+}\right)
    - \frac{\pi^2}{6} - \frac12\log^2w_+ \bigg]
  \nl
  &= -{q^2 \over 6M^2} + \order(Q^4) \,,
\end{align}
where in the last line, the result is expanded around $Q^2=2M^2(w-1) \to 0$.  

There are also several conventions in the atomic physics literature
for $\Delta K$, or equivalently for the proton electron and magnetic radii.
Let us define
\begin{align}\label{eq:relate}
  \frac16 r_{E}^2 &\equiv \frac{1}{ G_E(0)} {dG_E^{\rm Born}\over dq^2}\bigg|_{q^2=0} 
  \nl
  &= \tilde{F}_1^\prime(0) + \dfrac{F_2(0)}{4M^2}
  - {Z^2\alpha\over 2\pi M^2}\bigg[
    \frac23 \log{M^2\over \lambda^2} - M^2 \Delta K^\prime(0) \bigg] \,.
\end{align}
The case $\Delta K^\prime(0)=0$, as for $\Delta K^{\rm fac}$ in Eq.~(\ref{eq:Kfac}),
corresponds to the convention used in Ref.~\cite{Hill:2011wy}; in this convention,
the charge radius of a point particle vanishes including $\order(\alpha)$ radiative
corrections.
With the convention (\ref{eq:LAH}), we have instead
\be
(r_E^2)^{\rm LAH} = (r_E^2)^{\rm fac.} -{Z^2\alpha \over 2\pi M^2} \,. 
\ee
Several other conventions have been used, e.g. Pachucki's
definition in Ref.~\cite{Pachucki:1999zza}
implies
\be
(r_E^2)^{\rm P} = (r_E^2)^{\rm fac.} -{5Z^2\alpha \over 3\pi M^2} \,. 
\ee
Formula (\ref{eq:relate}) may be used to translate the radius used in other
conventions.  

\section{Phase space integrals \label{sec:phase}}

We list here expressions for phase space integrals used in the paper.
In terms of arbitrary timelike unit vectors $v^\mu$ and $v^{\prime \mu}$,~\cite{'tHooft:1978xw}
\begin{align}\label{eq:phasevvp} 
  &\int_{\ell^0 \le \Delta E} {d^3\ell \over (2\pi)^3  2\ell^0}
  {1 \over (v\cdot \ell)(v^\prime \cdot \ell) }
  =
  {1\over 8\pi^2  \sqrt{w^2-1}}\bigg[
    2\log(w_+) \log{ 2\Delta E\over \lambda}
 +    \log^2(v^0_+) -\log^2(v^{\prime 0}_+)
 \nl
 &\quad 
 + {\rm Li}_2\left( 1 - { v^0_+ \over \sqrt{w^2-1}} ( w_+ v^0 - v^{\prime 0} ) \right)
+ {\rm Li}_2\left( 1 - { v^0_- \over \sqrt{w^2-1}} ( w_+ v^0 - v^{\prime 0} ) \right)
\nl
&\quad
- {\rm Li}_2\left( 1 - { v^{\prime 0}_+ \over \sqrt{w^2-1}} ( v^0 - w_- v^{\prime 0} ) \right)
- {\rm Li}_2\left( 1 - { v^{\prime 0}_- \over \sqrt{w^2-1}} ( v^0 - w_- v^{\prime 0} ) \right)
\bigg] 
\,. 
\end{align}
In the limit $v^\prime = v$, Eq.~(\ref{eq:phasevvp}) becomes 
\begin{align}
  &\int_{\ell^0 \le \Delta E} {d^3\ell \over (2\pi)^3  2\ell^0}
  {1 \over (v\cdot \ell)^2}
  = {1\over 4\pi^2}\bigg[ \log{2\Delta E \over \lambda} - {v^0\over\sqrt{(v^0)^2-1}}\log(v^0 + \sqrt{(v^0)^2-1})
    \bigg] \,. 
\end{align}

In the analysis of power corrections (in $\Delta E/E$), we encounter 
integrals with the replacement $p^\prime \to \tilde{p}^{\prime\mu}$,
where $\tilde{p}^{\prime\mu}$ is defined with energy $\tilde{E}^\prime = E^\prime-\ell^0$
(recall $E^\prime=E$ for the static limit) and spatial momentum in the direction identical
to $p^{\prime\mu}$.  The first class of integrals is unchanged, 
\begin{align}
  &\int_{\ell^0 \le \Delta E} {d^3\ell \over (2\pi)^3  2\ell^0}
  {m^2 \over (\tilde{p}^\prime \cdot \ell)^2}
  =
\int_{\ell^0 \le \Delta E} {d^3\ell \over (2\pi)^3  2\ell^0}
    {m^2 \over ( p^\prime \cdot \ell)^2}
\to {1\over 4\pi^2} \bigg[\log{\Delta E\over E} - \log{\lambda\over m} \bigg] 
    \,,
\end{align}
where the arrow indicates the large energy limit, $v^0=E/m \to \infty$.  
For the second class of integrals, 
\begin{align}
  &\int_{\ell^0 \le \Delta E} {d^3\ell \over (2\pi)^3  2\ell^0}
  { 2p\cdot \tilde{p}^\prime  \over (p\cdot \ell)(\tilde{p}^\prime \cdot \ell) }
  \to
     {1\over 8\pi^2}\bigg[ 4\bigg( \log{\Delta E \over E}+\log{\lambda\over m}\bigg)L
       +L^2 +2{\rm Li}_2\left(\cos^2{\theta\over 2}\right) - {2\pi^2\over 3}
     \nl
     &\quad
     {\Delta E \over E} \left( -8 L + 4 \right)  + \dots \bigg] \,,
\end{align}
where the first order power correction is displayed. 

\section{Two loop mixed real-virtual correction: full theory \label{sec:2loopfull} }

Here we give details on the explicit evaluation of the two-loop matching
calculation for electron-proton scattering involving mixed real-virtual
corrections in the static source limit.  Recall the tree level squared matrix
element for the process without photon emission, 
\be
\sum |{\cal M}_0|^2 = e^2 {\rm Tr}[ (\slash{p}^\prime + m)\gamma^0 (\slash{p}+m)\gamma^0 ] \,. 
\ee
The squared matrix element for the process with photon emission is 
\be\label{eq:M1tree}
\sum |{\cal M}_{1,{\rm tree}}|^2 = \sum |{\cal M}_0|^2 e^2
\bigg[ {2p\cdot p^\prime \over p\cdot\ell p^\prime\cdot\ell} - {m^2\over (p\cdot \ell)^2}
  - {m^2\over (p^\prime\cdot\ell)^2} \bigg]  \,,
\ee
where terms yielding power suppressed
contributions after photon phase space integration have been dropped.

In the analysis of the phase space integrals for loop corrections to Eq.~(\ref{eq:M1tree}),
we encounter integrals of the form
\be\label{eq:I123}
I_1 = \int{d^3\ell \over \ell^0} {m^2\over (p\cdot \ell)^2} f(a) \,,
\quad I_2 = \int{d^3\ell \over \ell^0} {m^2\over (p^\prime\cdot\ell)^2} g(b)  \,,
\quad
I_3 = \int{d^3\ell \over \ell^0} {2p\cdot p^\prime \over p\cdot\ell p^\prime\cdot\ell } h(a,b)  \,,
\ee
where we introduce the shorthand $a=-p^\prime\cdot\ell/m^2-i0$, $b=p\cdot\ell/m^2$.
Introduce the small parameter $\kappa=m/E$.
For simplicity in this description, consider the case of backward scattering
where $\bm{p}^\prime = -\bm{p}$.  Introduce a light-cone basis for the photon momentum,
\be
\ell^\mu = (n\cdot \ell\, \bar{n}\cdot \ell, \ell_\perp^\mu) \,, 
\ee
where $n$ and $\bar{n}$ are lightlike vectors in the direction of $p$ and $p^\prime$, with 
$n^2=\bar{n}^2=0$, $n\cdot \bar{n}=2$.  For $I_1$, the leading contribution is
readily found to be 
\begin{align}\label{eq:Alead}
  k^\mu \sim (\kappa,\kappa^3,\kappa^2):\quad I_1&\sim f(\kappa) \to f(0) \,, 
\end{align}
i.e., from photons that are both soft and collinear to the final state electron. 
Contributions from other regions involve power suppression, e.g.
\begin{align}
  k^\mu \sim (\kappa,\kappa,\kappa):\quad  I_1&\sim \kappa^2 f(\kappa^{-1}) \,,
  \nl
  k^\mu \sim (\kappa^2,\kappa^2,\kappa^2):\quad  I_1&\sim \kappa^2 f(\kappa^0) \,. 
\end{align}
Similarly, for $I_2$, the leading contribution is from photons that are
both soft and collinear to the initial state electron, 
\begin{align}\label{eq:Blead}
  k^\mu \sim (\kappa^3,\kappa,\kappa^2):\quad I_2&\sim g(\kappa) \to g(0) \,. 
\end{align}
Finally, for $I_3$, multiple regions potentially contribute.
\begin{align}\label{eq:Clead}
  k^\mu \sim (\kappa,\kappa,\kappa):\quad I_3&\sim h(\kappa^{-1},\kappa^{-1} ) \,,
  \nl
  k^\mu \sim (\kappa^3,\kappa,\kappa^2):\quad I_3&\sim h(\kappa,\kappa^{-1} ) \,,  
  \nl
  k^\mu \sim (\kappa,\kappa^3,\kappa^2):\quad I_3&\sim h(\kappa^{-1},\kappa ) \,,  
  \nl
  k^\mu \sim (\kappa^2,\kappa^2,\kappa^2):\quad I_3&\sim h(\kappa^0,\kappa^{0} ) \,.
\end{align}
Inside loops, the presence of multiple momentum modes of the same virtuality ($L^2\sim \kappa^4$)
complicates a simple argument for soft-collinear factorization based on eikonal
decoupling (cf. the discussion surrounding Eq.~(\ref{eq:decouple}), where only a single,
soft, momentum mode is present).%
\footnote{
For a related discussion on potential difficulties with naive factorization, see Ref.~\cite{Becher:2007cu}. 
}
We proceed by direct evaluation of the diagrams. 

The relevant squared matrix element contains interference terms
between the tree-level real radiation diagrams of Fig.~\ref{fig:proton} and the
one loop real radiation diagrams of Fig.~\ref{fig:first}.  After averaging and
summing over initial and final electron spins, the squared matrix element,
divided by the tree level squared matrix element without radiation,
can be expanded in terms of the following basic integrals
(and the integrals related by $p \leftrightarrow p^\prime$, $\ell \leftrightarrow -\ell$), 
\begin{multline}\label{eq:ints}
  \int {1\over D_1(\lambda) D_2 D_3 D_4} \,, \quad
  \int {1\over D_1 D_2 D_3 D_4 } [1,L^\mu,L^\mu L^\nu, L^\mu L^\nu L^\rho] \,, \quad
  \int {1\over D_1 D_2 D_4} [1,L^\mu,L^\mu L^\nu]  \,,
  \\
  \int {1\over D_1 D_3 D_4} [1,L^\mu,L^\mu L^\nu]  \,, \quad
  \int {1\over D_1 D_4} [1,L^\mu] \,, 
\end{multline}
where integration is over $\int = \int d^dL$, and the denominators are  
\begin{multline}
  D_1(\lambda) = L^2-\lambda^2 \,, \quad
  D_1 = L^2 \,, \quad
  D_2 =  L^2+2L\cdot p \,, \quad
  D_3 =  L^2+2L\cdot p^\prime \,, \quad
  \\
  D_4 = L^2+2L\cdot(p^\prime+\ell)+2p^\prime\cdot \ell \,. 
\end{multline}
We evaluated these integrals
using dimensional regularization for ultraviolet divergences and photon
mass $\lambda$ for infrared divergences.   After mass, coupling and wavefunction renormalization, and
expressing the result in terms of the onshell coupling, 
we obtain expressions of the form (\ref{eq:I123}), which may be
expanded according to Eqs.~(\ref{eq:Alead}), (\ref{eq:Blead}) and (\ref{eq:Clead}). 
Neglecting contributions that are power suppressed after photon phase space integration, 
the final result reads
\begin{align}\label{eq:direct}
\sum |{\cal M}_{1}|^2 &= \sum |{\cal M}_0|^2 e^2
\bigg[ {2p\cdot p^\prime \over p\cdot\ell p^\prime\cdot\ell } - {m^2\over (p\cdot\ell)^2} - {m^2\over (p^\prime\cdot\ell)^2} \bigg]
\bigg\{ 1 + {\alpha\over 4\pi}\bigg[ -2\log^2{Q^2\over m^2} \nl
  &\quad + 8\log{\lambda\over m}\left(\log{Q^2\over m^2} -1\right)
  + 6\log{Q^2\over m^2} + {2\pi^2\over 3} - 8\bigg] \bigg\}
\,.
\end{align}

\section{Two loop mixed real-virtual correction: effective theory \label{sec:2loopeff} }

\begin{figure}[t]
  \begin{center}
\begin{align}
    \parbox{30mm}{
      \begin{fmfgraph*}(70,40)
        \fmfset{arrow_len}{3mm}\fmfset{arrow_ang}{12}
        \fmfleftn{l}{3}
        \fmfrightn{r}{3}
        \fmfbottomn{b}{1}
        \fmf{phantom}{l3,x,v,y,r3}
        \fmf{photon, tension=0.6}{b1,v}
        \fmffreeze
        \fmf{fermion}{l3,x}
        \fmf{fermion}{x,v,y}
        \fmf{fermion}{y,r3}
        \fmf{photon,left}{x,y}
        \fmfv{d.sh=circle,d.f=empty,d.si=.1w,l=$\times$,label.dist=0}{b1}
      \end{fmfgraph*}
    }
    &=
    \parbox{30mm}{
      \begin{fmfgraph*}(70,40)
        \fmfset{arrow_len}{3mm}\fmfset{arrow_ang}{12}
        \fmfleftn{l}{3}
        \fmfrightn{r}{3}
        \fmfbottomn{b}{1}
        \fmf{phantom}{l3,x,v,y,r3}
        \fmf{photon, tension=0.6}{b1,v}
        \fmffreeze
        \fmf{fermion}{l3,v,r3}
        \fmfv{d.sh=circle,d.f=empty,d.si=.1w,l=$\times$,label.dist=0}{b1}
        \fmfv{d.sh=circle,d.f=full,d.si=.1w}{v}
      \end{fmfgraph*}
    }
    +
    \parbox{30mm}{
      \begin{fmfgraph*}(70,40)
        \fmfset{arrow_len}{3mm}\fmfset{arrow_ang}{12}
        \fmfleftn{l}{3}
        \fmfrightn{r}{3}
        \fmfbottomn{b}{1}
        \fmf{phantom}{l3,x,v,y,r3}
        \fmf{photon, tension=0.6}{b1,v}
        \fmffreeze
        \fmf{fermion}{l3,x,v,y,r3}
        \fmf{curly,left}{x,y}
        \fmfv{d.sh=circle,d.f=empty,d.si=.1w,l=$\times$,label.dist=0}{b1}
      \end{fmfgraph*}
    }
    +
         \parbox{30mm}{
      \begin{fmfgraph*}(70,40)
        \fmfset{arrow_len}{3mm}\fmfset{arrow_ang}{12}
        \fmfleftn{l}{3}
        \fmfrightn{r}{3}
        \fmfbottomn{b}{1}
        \fmf{phantom}{l3,x,v,y,r3}
        \fmf{photon, tension=0.6}{b1,v}
        \fmffreeze
        \fmf{fermion}{l3,v,y,r3}
        \fmf{curly,left=2}{v,y}
        \fmf{plain,left=2}{v,y}
        \fmfv{d.sh=circle,d.f=empty,d.si=.1w,l=$\times$,label.dist=0}{b1}
      \end{fmfgraph*}
         }
             +
         \parbox{30mm}{
      \begin{fmfgraph*}(70,40)
        \fmfset{arrow_len}{3mm}\fmfset{arrow_ang}{12}
        \fmfleftn{l}{3}
        \fmfrightn{r}{3}
        \fmfbottomn{b}{1}
        \fmf{phantom}{l3,x,v,y,r3}
        \fmf{photon, tension=0.6}{b1,v}
        \fmffreeze
        \fmf{fermion}{l3,x,v,r3}
        \fmf{curly,left=2}{x,v}
        \fmf{plain,left=2}{x,v}
        \fmfv{d.sh=circle,d.f=empty,d.si=.1w,l=$\times$,label.dist=0}{b1}
      \end{fmfgraph*}
         }
         \nonumber
\end{align}
\caption{Expansion in momentum regions of amplitudes for electron scattering in the static source limit.
  Diagram on the left hand side is in the full theory (QED), diagrams on the right hand side are in the  effective theory.
  Soft and collinear photons are represented by curly lines, and curly lines superimposed on solid lines,
  respectively. 
  \label{fig:scet}
  }
  \end{center}
\end{figure}
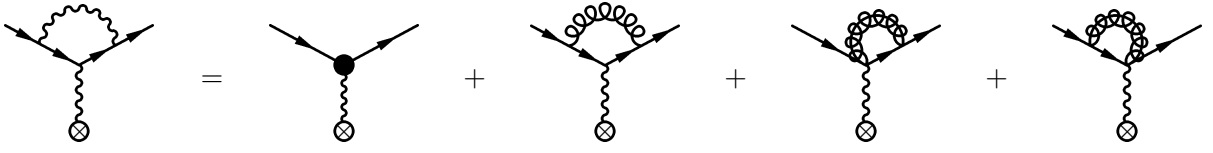

\begin{figure}[t]
  \begin{center}
    \begin{align}
      a:\quad
      \parbox{30mm}{
        \begin{fmfgraph*}(80,40)
          \fmfleftn{l}{4}
          \fmfrightn{r}{4}
          \fmfbottomn{b}{1}
          \fmf{phantom}{l4,x1,x2,v,y1,y2,r4}
          \fmf{photon, tension=0.6}{b1,v}
          \fmffreeze
          \fmf{fermion}{l4,x1,v,y1,y2,r4}
          \fmf{phantom,tag=1}{y1,r2}
          \fmfipath{p[]}
          \fmfiset{p1}{vpath1(__y1,__r2)}
          \fmfi{photon}{ point 0length(p1) of p1 -- point length(p1)/2 of p1}
          \fmf{photon,left=0.5}{x1,y2}
          \fmfv{d.sh=circle,d.f=empty,d.si=.1w,l=$\times$,label.dist=0}{b1}
        \end{fmfgraph*}
      }
      &=
      \parbox{30mm}{
        \begin{fmfgraph*}(80,40)
          \fmfleftn{l}{4}
          \fmfrightn{r}{4}
          \fmfbottomn{b}{1}
          \fmf{phantom}{l4,x1,x2,v,y1,y2,r4}
          \fmf{photon, tension=0.6}{b1,v}
          \fmffreeze
          \fmf{fermion}{l4,x1,v,y1,y2,r4}
          \fmf{phantom,tag=1}{y1,r2}
          \fmfipath{p[]}
          \fmfiset{p1}{vpath1(__y1,__r2)}
          \fmfi{curly}{ point 0length(p1) of p1 -- point 3length(p1)/4 of p1}
          \fmf{curly,left=0.5}{x1,y2}
          \fmfv{d.sh=circle,d.f=empty,d.si=.1w,l=$\times$,label.dist=0}{b1}
        \end{fmfgraph*}
      }
      \parbox{30mm}{
        \begin{fmfgraph*}(80,40)
          \fmfleftn{l}{4}
          \fmfrightn{r}{4}
          \fmfbottomn{b}{1}
          \fmf{phantom}{l4,x1,x2,v,y1,y2,r4}
          \fmf{photon, tension=0.6}{b1,v}
          \fmffreeze
          \fmf{fermion}{l4,x1,v,y1,y2,r4}
          \fmf{phantom,tag=1}{y1,r2}
          \fmfipath{p[]}
          \fmfiset{p1}{vpath1(__y1,__r2)}
          \fmfi{curly}{ point 0length(p1) of p1 -- point 3length(p1)/4 of p1}
          \fmf{curly,left=1}{v,y2}
          \fmf{plain,left=1}{v,y2}
          \fmfv{d.sh=circle,d.f=empty,d.si=.1w,l=$\times$,label.dist=0}{b1}
        \end{fmfgraph*}
      }
      \nonumber
      \\[5mm]
      b:\quad
      \parbox{30mm}{
        \begin{fmfgraph*}(80,40)
          \fmfset{arrow_len}{3mm}\fmfset{arrow_ang}{12}
          \fmfleftn{l}{5}
          \fmfrightn{r}{5}
          \fmfbottomn{b}{1}
          \fmf{phantom}{l5,x1,x2,v,y1,y2,r5}
          \fmf{photon, tension=0.6}{b1,v}
          \fmffreeze
          \fmf{fermion}{l5,x2,v,y1,y2,r5}
          \fmf{phantom,tag=1}{y2,r3}
          \fmfipath{p[]}
          \fmfiset{p1}{vpath1(__y2,__r3)}
          \fmfi{photon}{ point 0length(p1) of p1 -- point 2length(p1)/3 of p1}
          \fmf{photon,left=0.5}{x2,y1}
          \fmfv{d.sh=circle,d.f=empty,d.si=.1w,l=$\times$,label.dist=0}{b1}
        \end{fmfgraph*}
      }
      &=
      \parbox{30mm}{
        \begin{fmfgraph*}(80,40)
          \fmfset{arrow_len}{3mm}\fmfset{arrow_ang}{12}
          \fmfleftn{l}{5}
          \fmfrightn{r}{5}
          \fmfbottomn{b}{1}
          \fmf{phantom}{l5,x1,x2,v,y1,y2,r5}
          \fmf{photon, tension=0.6}{b1,v}
          \fmffreeze
          \fmf{fermion}{l5,v,y2,r5}
          \fmf{phantom,tag=1}{y2,r3}
          \fmfipath{p[]}
          \fmfiset{p1}{vpath1(__y2,__r3)}
          \fmfi{photon}{ point 0length(p1) of p1 -- point 2length(p1)/3 of p1}
          \fmfv{d.sh=circle,d.f=empty,d.si=.1w,l=$\times$,label.dist=0}{b1}
          \fmfv{d.sh=circle,d.f=full,d.si=.1w}{v}
        \end{fmfgraph*}
      }
      +
      \parbox{30mm}{
        \begin{fmfgraph*}(80,40)
          \fmfset{arrow_len}{3mm}\fmfset{arrow_ang}{12}
          \fmfleftn{l}{5}
          \fmfrightn{r}{5}
          \fmfbottomn{b}{1}
          \fmf{phantom}{l5,x1,x2,v,y1,y2,r5}
          \fmf{photon, tension=0.6}{b1,v}
          \fmffreeze
          \fmf{fermion}{l5,x2,v,y1,y2,r5}
          \fmf{phantom,tag=1}{y2,r3}
          \fmfipath{p[]}
          \fmfiset{p1}{vpath1(__y2,__r3)}
          \fmfi{curly}{ point 0length(p1) of p1 -- point 2length(p1)/3 of p1}
          \fmf{curly,left=1}{x2,y1}
          \fmfv{d.sh=circle,d.f=empty,d.si=.1w,l=$\times$,label.dist=0}{b1}
        \end{fmfgraph*}
      }
      + 
      \parbox{30mm}{
        \begin{fmfgraph*}(80,40)
          \fmfset{arrow_len}{3mm}\fmfset{arrow_ang}{12}
          \fmfleftn{l}{5}
          \fmfrightn{r}{5}
          \fmfbottomn{b}{1}
          \fmf{phantom}{l5,x1,x2,v,y1,y2,r5}
          \fmf{photon, tension=0.6}{b1,v}
          \fmffreeze
          \fmf{fermion}{l5,v,y1,y2,r5}
          \fmf{phantom,tag=1}{y2,r3}
          \fmfipath{p[]}
          \fmfiset{p1}{vpath1(__y2,__r3)}
          \fmfi{curly}{ point 0length(p1) of p1 -- point 2length(p1)/3 of p1}
          \fmf{curly,left=2}{v,y1}
          \fmf{plain,left=2}{v,y1}
          \fmfv{d.sh=circle,d.f=empty,d.si=.1w,l=$\times$,label.dist=0}{b1}
        \end{fmfgraph*}
      }
      \nonumber
      \\[5mm]
      &\quad
      + 
      \parbox{30mm}{
        \begin{fmfgraph*}(80,40)
          \fmfset{arrow_len}{3mm}\fmfset{arrow_ang}{12}
          \fmfleftn{l}{5}
          \fmfrightn{r}{5}
          \fmfbottomn{b}{1}
          \fmf{phantom}{l5,x1,x2,v,y1,y2,r5}
          \fmf{photon, tension=0.6}{b1,v}
          \fmffreeze
          \fmf{fermion}{l5,x2,v,y2,r5}
          \fmf{phantom,tag=1}{y2,r3}
          \fmfipath{p[]}
          \fmfiset{p1}{vpath1(__y2,__r3)}
          \fmfi{curly}{ point 0length(p1) of p1 -- point 2length(p1)/3 of p1}
          \fmf{curly,left=2}{x2,v}
          \fmf{plain,left=2}{x2,v}
          \fmfv{d.sh=circle,d.f=empty,d.si=.1w,l=$\times$,label.dist=0}{b1}
        \end{fmfgraph*}
      }
      \nonumber
      \\[5mm]
      c:\quad
      \parbox{30mm}{
        \begin{fmfgraph*}(80,40)
          \fmfset{arrow_len}{3mm}\fmfset{arrow_ang}{12}
          \fmfleftn{l}{4}
          \fmfrightn{r}{4}
          \fmfbottomn{b}{1}
          \fmf{phantom}{l4,x1,x2,v,y1,y2,y3,r4}
          \fmf{photon, tension=0.6}{b1,v}
          \fmffreeze
          \fmf{fermion}{l4,v,y1,y2,y3,r4}
          \fmf{phantom,tag=1}{y2,r2}
          \fmfipath{p[]}
          \fmfiset{p1}{vpath1(__y2,__r2)}
          \fmfi{photon}{ point 0length(p1) of p1 -- point 2length(p1)/3 of p1}
          \fmf{photon,left}{y1,y3}
          \fmfv{d.sh=circle,d.f=empty,d.si=.1w,l=$\times$,label.dist=0}{b1}
        \end{fmfgraph*}
      }
      &=
      \parbox{30mm}{
        \begin{fmfgraph*}(80,40)
          \fmfset{arrow_len}{3mm}\fmfset{arrow_ang}{12}
          \fmfleftn{l}{4}
          \fmfrightn{r}{4}
          \fmfbottomn{b}{1}
          \fmf{phantom}{l4,x1,x2,v,y1,y2,y3,r4}
          \fmf{photon, tension=0.6}{b1,v}
          \fmffreeze
          \fmf{fermion}{l4,v,y1,y2,y3,r4}
          \fmf{phantom,tag=1}{y2,r2}
          \fmfipath{p[]}
          \fmfiset{p1}{vpath1(__y2,__r2)}
          \fmfi{curly}{ point 0length(p1) of p1 -- point 2length(p1)/3 of p1}
          \fmf{curly,left=1.5}{y1,y3}
          \fmfv{d.sh=circle,d.f=empty,d.si=.1w,l=$\times$,label.dist=0}{b1}
        \end{fmfgraph*}
      }
      +
      \parbox{30mm}{
        \begin{fmfgraph*}(80,40)
          \fmfset{arrow_len}{3mm}\fmfset{arrow_ang}{12}
          \fmfleftn{l}{4}
          \fmfrightn{r}{4}
          \fmfbottomn{b}{1}
          \fmf{phantom}{l4,x1,x2,v,y1,y2,y3,r4}
          \fmf{photon, tension=0.6}{b1,v}
          \fmffreeze
          \fmf{fermion}{l4,v,y1,y2,y3,r4}
          \fmf{phantom,tag=1}{y2,r2}
          \fmfipath{p[]}
          \fmfiset{p1}{vpath1(__y2,__r2)}
          \fmfi{curly}{ point 0length(p1) of p1 -- point 2length(p1)/3 of p1}
          \fmf{curly,left=1.5}{y1,y3}
          \fmf{plain,left=1.5}{y1,y3}
          \fmfv{d.sh=circle,d.f=empty,d.si=.1w,l=$\times$,label.dist=0}{b1}
        \end{fmfgraph*}
      }
      \nonumber
      \\[5mm]
      d:\quad
      \parbox{30mm}{
        \begin{fmfgraph*}(80,40)
          \fmfset{arrow_len}{3mm}\fmfset{arrow_ang}{12} 
          \fmfleftn{l}{5}
          \fmfrightn{r}{5}
          \fmfbottomn{b}{1}
          \fmf{phantom}{l5,x1,x2,v,y1,y2,y3,r5}
          \fmf{photon, tension=0.6}{b1,v}
          \fmffreeze
          \fmf{fermion}{l5,v,y1,y2,y3,r5}
          \fmf{phantom,tag=1}{y3,r3}
          \fmfipath{p[]}
          \fmfiset{p1}{vpath1(__y3,__r3)}
          \fmfi{photon}{ point 0length(p1) of p1 -- point 2length(p1)/3 of p1}
          \fmf{photon,tension=0.5, left=2}{y1,y2}
          \fmfv{d.sh=circle,d.f=empty,d.si=.1w,l=$\times$,label.dist=0}{b1}
        \end{fmfgraph*}
      }
      &=
      \parbox{30mm}{
        \begin{fmfgraph*}(80,40)
          \fmfset{arrow_len}{3mm}\fmfset{arrow_ang}{12} 
          \fmfleftn{l}{5}
          \fmfrightn{r}{5}
          \fmfbottomn{b}{1}
          \fmf{phantom}{l5,x1,x2,v,y1,y2,y3,r5}
          \fmf{photon, tension=0.6}{b1,v}
          \fmffreeze
          \fmf{fermion}{l5,v,y1,y2,y3,r5}
          \fmf{phantom,tag=1}{y3,r3}
          \fmfipath{p[]}
          \fmfiset{p1}{vpath1(__y3,__r3)}
          \fmfi{curly}{ point 0length(p1) of p1 -- point 2length(p1)/3 of p1}
          \fmf{curly,tension=0.5, left=2}{y1,y2}
          \fmfv{d.sh=circle,d.f=empty,d.si=.1w,l=$\times$,label.dist=0}{b1}
        \end{fmfgraph*}
      }
      +
      \parbox{30mm}{
        \begin{fmfgraph*}(80,40)
          \fmfset{arrow_len}{3mm}\fmfset{arrow_ang}{12} 
          \fmfleftn{l}{5}
          \fmfrightn{r}{5}
          \fmfbottomn{b}{1}
          \fmf{phantom}{l5,x1,x2,v,y1,y2,y3,r5}
          \fmf{photon, tension=0.6}{b1,v}
          \fmffreeze
          \fmf{fermion}{l5,v,y1,y2,y3,r5}
          \fmf{phantom,tag=1}{y3,r3}
          \fmfipath{p[]}
          \fmfiset{p1}{vpath1(__y3,__r3)}
          \fmfi{curly}{ point 0length(p1) of p1 -- point 2length(p1)/3 of p1}
          \fmf{curly, tension=0.5, left=2}{y1,y2}
          \fmf{plain,tension=0.5, left=2}{y1,y2}
          \fmfv{d.sh=circle,d.f=empty,d.si=.1w,l=$\times$,label.dist=0}{b1}
        \end{fmfgraph*}
      }
      \nonumber
      \\[5mm]
      e:\quad
      \parbox{30mm}{
        \begin{fmfgraph*}(60,40)
          \fmfset{arrow_len}{3mm}\fmfset{arrow_ang}{12} 
          \fmfleftn{l}{3}
          \fmfrightn{r}{3}
          \fmf{phantom}{l2,x,v,y,r2}
          \fmffreeze
          \fmf{fermion}{l2,x,y,r2}
          \fmf{photon,left}{x,y}
        \end{fmfgraph*}
      }
      &=
      \parbox{30mm}{
        \begin{fmfgraph*}(60,40)
          \fmfset{arrow_len}{3mm}\fmfset{arrow_ang}{12} 
          \fmfleftn{l}{3}
          \fmfrightn{r}{3}
          \fmf{phantom}{l2,x,v,y,r2}
          \fmffreeze
          \fmf{fermion}{l2,x,y,r2}
          \fmf{curly,left}{x,y}
      \end{fmfgraph*}
      }
      +
      \parbox{30mm}{
        \begin{fmfgraph*}(60,40)
          \fmfset{arrow_len}{3mm}\fmfset{arrow_ang}{12} 
          \fmfleftn{l}{3}
          \fmfrightn{r}{3}
          \fmf{phantom}{l2,x,v,y,r2}
          \fmffreeze
          \fmf{fermion}{l2,x,y,r2}
          \fmf{curly,left}{x,y}
          \fmf{plain,left}{x,y}
      \end{fmfgraph*}
      }
      \nonumber
    \end{align}    
    \caption{Same as Fig.~\ref{fig:scet}, but for electron scattering with real photon emission. 
      \label{fig:scet1}
    }
  \end{center}
\end{figure}
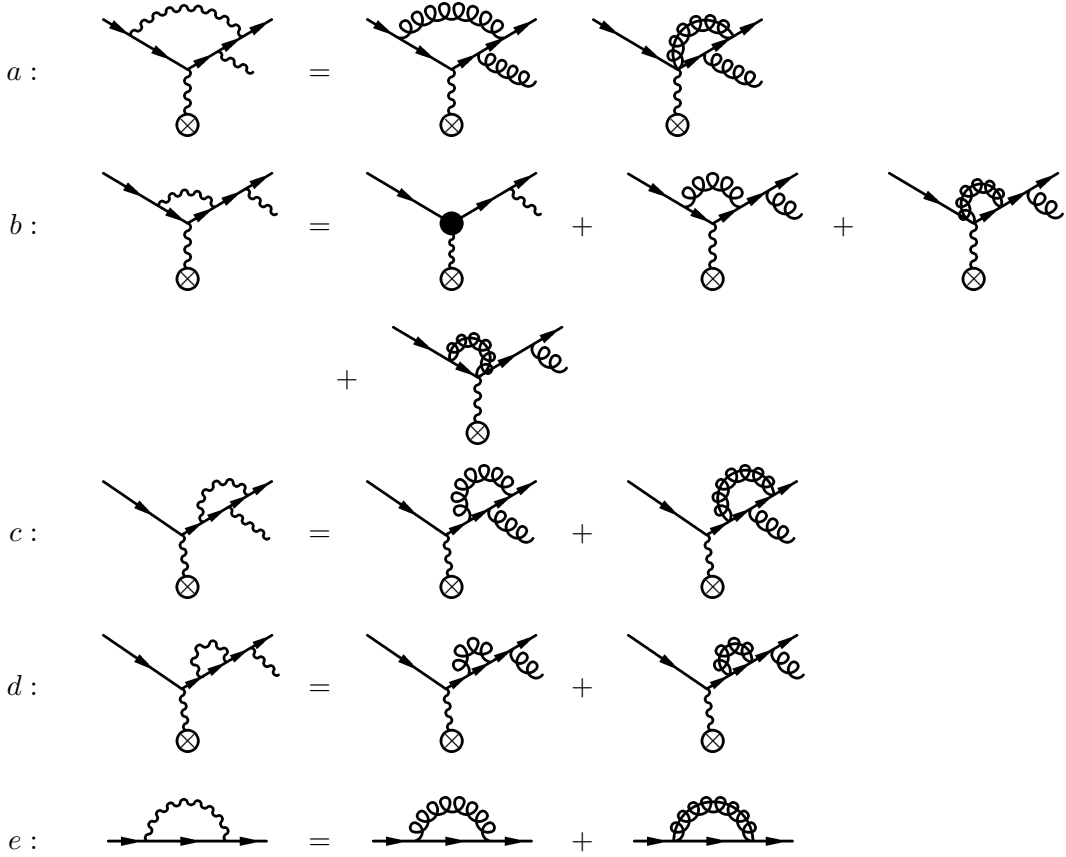

Here we outline the evaluation of the mixed real-virtual corrections
using a decomposition into soft and collinear momentum regions, formalized
as soft-collinear effective theory~\cite{Bauer:2000ew,Bauer:2000yr,Bauer:2001ct,Bauer:2001yt,Chay:2002vy,Beneke:2002ph,Hill:2002vw,Becher:2014oda}. 
We first review the analysis of vertex corrections. 

\subsection{Vertex corrections} 

Consider the amplitude pictured on the left hand side of Fig.~\ref{fig:scet},
\begin{align}
    \delta F \gamma^\mu &= -ie^2 \int {d^dL \over (2\pi)^d}
  \gamma^\alpha (\slash{L} + \slash{p}^\prime + m)\gamma^\mu (\slash{L}+\slash{p} + m ) \gamma_\alpha
  {1\over L^2-\lambda^2} {1\over L^2+2L\cdot p} {1\over L^2+2L\cdot p^\prime} \,, 
\end{align}
and the corresponding decomposition pictured on the right hand side of
Fig.~\ref{fig:scet}.
Introduce light-cone vectors $n^\mu$ and $\bar{n}^\mu$ for the direction $p^\mu$,
and corresponding vectors $n^{\prime \mu}$ and $\bar{n}^{\prime\mu}$ for the direction $p^{\prime\mu}$.
The hard contribution is represented by the first diagram on the right hand side of Fig.~\ref{fig:scet},
and is obtained from
\begin{align}
  \delta F_H \gamma^\mu &= -ie^2 \int {d^dL \over (2\pi)^d}
  \gamma^\alpha (\slash{L} + \slash{p}_{-^\prime}^\prime)\gamma^\mu (\slash{L}+\slash{p}_-) \gamma_\alpha
  {1\over L^2} {1\over L^2+2L\cdot p_-} {1\over L^2+2L\cdot p_{-^\prime}^\prime}
  \nl
  &= -ie^2 \gamma^\mu [c_\epsilon] Q^{-2\epsilon} \left[ -{2\over \epsilon^2} - {3\over \epsilon} - 8 + {\pi^2\over 3} \right] \,,
\end{align}
where $[c_\epsilon]\equiv i (4\pi)^{-2+\epsilon} \Gamma(1+\epsilon)$, 
and $p_-^\mu \equiv \bar{n}\cdot p\, n^\mu/2$ is the large component of the momentum $p^\mu$
(similarly $p^{\prime\mu}_{-^\prime}$ is defined in terms of $n^{\prime\mu}$ and $\bar{n}^{\prime\mu}$).  
This yields the one loop contribution to $F_H^{\rm bare}$ in Eq.~(\ref{eq:hard}).%
\footnote{
Recall that our definition of $\alpha_{\rm bare}$ absorbs $e^{-\gamma_E \epsilon}$, whereas
$[c_\epsilon]$ contains $\Gamma(1+\epsilon)=e^{-\gamma_E \epsilon}(1+ \epsilon^2 \pi^2/12 + \dots)$.
}

The soft contribution corresponds to the second diagram on the right hand side of Fig.~\ref{fig:scet},
\begin{align}
  \delta F_S\gamma^\mu &= -ie^2\gamma^\mu \int {d^dL \over (2\pi)^d} 4p\cdot p^\prime 
         {1\over L^2} {1 \over 2 L \cdot p} {1 \over 2 L\cdot p^\prime} 
  = -ie^2\gamma^\mu [c_\epsilon] \lambda^{-2\epsilon} \left[ -{2\over \epsilon} \log{Q^2\over m^2} \right] \,. 
\end{align} 
Combined with the soft contribution to onshell wavefunction renormalization [$Z_h$ in Eq.~(\ref{eq:Zonshell})],
this yields the one loop $F_S^{\rm bare}$ given by Eqs.~(\ref{eq:Sren}) and (\ref{eq:Smu}). 

The remaining contributions arise from momentum regions collinear to the final and initial
electron momenta, shown as the final diagrams on the right hand side of Fig.~\ref{fig:scet}.
The required basis of integrals is
\begin{align}
  [I_c,\,I_c^\mu,\,I_c^{\mu\nu}] &= \int{d^dL\over (2\pi)^d} {1\over L^2}{1\over 2L\cdot p_-}{1\over L^2+2L\cdot p^\prime}
  [1,\,L^\mu,\,L^\mu L^\nu] \,. 
\end{align}
We expand 
\begin{align}
  I_c &= [c_\epsilon]{1\over Q^2} I^{(0)} \,,
  \nl
  I_c^\mu &= [c_\epsilon]{1\over Q^2} \left[ I_1^{(1)} p_-^\mu + I_2^{(1)} p^{\prime\mu} \right] \,,
  \nl
  I_c^{\mu\nu} &= [c_\epsilon]\left[ g^{\mu\nu} I_1^{(2)}
    + {1\over Q^2} \left( I_2^{(2)} p_-^\mu p_-^\nu + I_3^{(2)}p^{\prime\mu}p^{\prime\nu}
    + I_4^{(2)} ( p_-^\mu p^{\prime\nu} + p_-^\nu p^{\prime\mu} )
    \right)
    \right] \,.  
\end{align}
Using these elementary integrals, we obtain 
\begin{align}
  \delta F_J &= -ie^2[c_\epsilon] \left[ 2I^{(0)} + 2 I^{(1)}_2 + (p\leftrightarrow p^\prime) \right]
  =
  -ie^2[c_\epsilon] m^{-2\epsilon} \left[ {2\over \epsilon^2} + {4\over \epsilon} + 8 \right]
  \,.
  \end{align} 
Combined with the collinear contribution to onshell wavefunction renormalization
[the difference of $Z_\Psi$ and $Z_h$ in Eq.~(\ref{eq:Zonshell})], this yields
the one loop $F_J^{\rm bare}$ given by Eq.~(\ref{eq:FJR12}) (recall that $F_R=1$ at one loop order).

The components of the factorization theorem (\ref{eq:facfull}) are thus identified
with effective theory contributions represented by the diagrams of Fig.~\ref{fig:scet}. 

\subsection{Real radiation} 

Consider now the case of real radiation at loop level.    
Begin with the interference between the diagram pictured in Fig.~\ref{fig:scet1}a, and the tree
level photon emission diagrams from Fig.~\ref{fig:proton}.   The relevant integrals
in the full theory evaluation are given by the first two terms of Eq.~(\ref{eq:ints}), with four
denominators.
Let us focus in particular on the scalar integral, 
\begin{align}
  I = \int {d^dL \over (2\pi)^d}
  {1\over L^2-\lambda^2}{1\over L^2+2L\cdot p}{1\over L^2+2L\cdot p^\prime}{1\over L^2+2L\cdot(p^\prime+k)+2p^\prime\cdot k} \,.
\end{align}
The soft photon contribution, represented by the first diagram on the RHS of Fig.~\ref{fig:scet1}a is
\begin{align}
  I_s &= \int {d^dL\over (2\pi)^d}  {1\over L^2-\lambda^2}{1\over 2L\cdot p}{1\over 2L\cdot p^\prime}{1\over 2(L+k)\cdot p^\prime}
  \nl
  &= {1\over Q^2}{1\over 2k\cdot p^\prime}[c_\epsilon] \bigg[
    -2\log^2{m^2\over Q^2} + 2\log{-2k\cdot p^\prime \over \lambda Q} \log{m^2\over Q^2} + {\pi^2\over 6}\bigg] \,. 
\end{align}
The collinear contribution,  represented by the second diagram on the RHS of Fig.~\ref{fig:scet1}a is
\begin{align}
  I_{c} &= \int {d^dL\over (2\pi)^d} {1\over L^2}{1\over 2 L_- \cdot p}{1\over L^2+2L\cdot p^\prime}{1\over L^2 + 2L\cdot p^\prime
    + 2 k_+ \cdot (L+p^\prime)}
\nl
&={1\over 2k\cdot p^\prime} \int (dL)
   {1\over L^2}\left( {1\over n \cdot p \bar{n} \cdot(L+p^\prime)} - {1\over n \cdot p \bar{n} \cdot L}
   \right)
   \nl
   &\quad \times 
   \left(
        {1\over L^2+ 2L\cdot p^\prime + n \cdot k \bar{n} \cdot (L+p^\prime)}
        - {1\over L^2 + 2L\cdot p^\prime}
        \right)  
        \nl
        &=  {1\over Q^2}{1\over 2k\cdot p^\prime}[c_\epsilon] {1\over \epsilon} J(1,0,0)
         = {1\over Q^2}{1\over 2k\cdot p^\prime}[c_\epsilon]
  (m^2a)^{-{2\epsilon}}m^{2\epsilon}\left( {1\over 2\epsilon^2} + {\pi^2\over 6} \right) \,, 
\end{align}
where we introduce the functions
\begin{align}
J(r,s,t) &= \int_0^1 dx x^s \left[ {1\over (1-x)^r} - {1\over (-x)^r} \right]
\left[ x(1-x)m^2a + x^2m^2)^{t-\epsilon} - (x^2m^2)^{t-\epsilon}  \right] \,. 
\end{align}
The presence of multiple low energy scales leads to a nontrivial subtraction in order
to avoid double counting.  The soft limit of the collinear integral is 
\begin{align}
  I_c\big|_{s} &= \int {d^dL\over (2\pi)^d} {1\over L^2}{1\over 2 L_- \cdot p}
  {1\over 2L\cdot p^\prime}
  {1\over 2(L+k_+) \cdot p^\prime }
  = {1\over Q^2}{1\over 2k\cdot p^\prime}[c_\epsilon]
  (m^2a)^{-{2\epsilon}}m^{2\epsilon}\left( -{1\over 2\epsilon^2} - {\pi^2\over 6} \right) \,, 
\end{align}
so that accounting for the overlap, the collinear region gives vanishing contribution,
\be
I_c - I_c\big|_{s} = 0 \,. 
\ee
The remaining integrals may be treated similarly.  For example,
consider 
\begin{align}
  I^\mu &= \int {d^dL \over (2\pi)^d}
  {1\over L^2}{1\over L^2+2L\cdot p}{1\over L^2+2L\cdot p^\prime}{1\over L^2+2L\cdot(p^\prime+k)+2p^\prime\cdot k}
  L^\mu \,. 
\end{align}
In the collinear region, we expand as
\begin{align}
  I^\mu_c &= \int {d^dL \over (2\pi)^d}
    {1\over L^2}{1\over 2L_-\cdot p}{1\over L^2+2L\cdot p^\prime}{1\over L^2+2L\cdot p^\prime + 2k_+\cdot(L+p^\prime)}
  L^\mu
  \nl
  &= {1\over Q^2}{1\over 2k\cdot p^\prime}[c_\epsilon] \left( I^{(1)}_1 p^{\prime \mu} + I^{(1)}_2 k_+^\mu \right) \,,
\end{align}
with
\begin{align}
  I^{(1)}_1 &=  -{1\over\epsilon} J(1,1,0) \,,
  \nl
  I^{(2)}_1 &= -{1\over\epsilon} K(1,1,0) + {1\over m^2a}{1\over \epsilon(1-\epsilon)} J(2,1,0) \,,
\end{align}
where $J(n,m,p)$ is given above and
\begin{align}
  K(r,s,t)
    &= \int_0^1 dx x^s \left[ {1\over (1-x)^r} - {1\over (-x)^r} \right] ( x(1-x)m^2a + x^2m^2)^{t-\epsilon} \,.
\end{align}
Explicit evaluation gives
\begin{align}
I^{(1)}_1 &= -{\rm Li}_2(1-a) + {\pi^2\over 6} \,. 
\end{align}
Similarly, consider  
\begin{align}
  I^{\mu\nu} &= \int {d^dL \over (2\pi)^d}
  {1\over L^2}{1\over L^2+2L\cdot p}{1\over L^2+2L\cdot p^\prime}{1\over L^2+2L\cdot(p^\prime+k)+2p^\prime\cdot k}
  L^\mu L^\nu \,.
\end{align}
In the collinear region, we expand as
\begin{align}
  I^{\mu\nu}_c &= \int {d^dL \over (2\pi)^d}
    {1\over L^2}{1\over 2L_-\cdot p}{1\over L^2+2L\cdot p^\prime}{1\over L^2+2L\cdot p^\prime + 2k_+\cdot(L+p^\prime)}
  L^\mu L^\nu 
  \nl
  &= {1\over Q^2}{1\over 2k\cdot p^\prime}[c_\epsilon] \left( I^{(2)}_1 g^{\mu\nu}
  + I^{(2)}_2 p^{\prime\mu}p^{\prime\nu}
  + I^{(2)}_3 (p^{\prime\mu} k_+^\nu + k_+^\mu p^{\prime \nu})
  + I^{(2)}_4 k_+^\mu k_+^\nu
  \right) \,,
\end{align}
with
\begin{align}
  I^{(2)}_1 &= {1\over 2\epsilon(1-\epsilon)} J(1,0,1)\,,
  \nl
  I^{(2)}_2 &= {1\over \epsilon} J(1,2,0) \,,
  \nl
  I^{(2)}_3 &= {1\over \epsilon} K(1,2,0) - {1\over m^2a}{1\over \epsilon(1-\epsilon)} J(2,1,1) \,,
  \nl
  I^{(2)}_4 &= {1\over \epsilon} K(1,2,0) - {2\over m^2a}{1\over \epsilon(1-\epsilon)} K(2,1,1) \,,
  \nl
  I^{(2)}_4 &= {1\over\epsilon} K(1,2,0) - {2\over m^2a}{1\over \epsilon(1-\epsilon)} K(2,1,1)
  + {2\over (m^2a)^2}{1\over \epsilon(1-\epsilon)(2-\epsilon)}J(3,0,2) \,. 
\end{align}
The relevant integrals are, explicitly,
\begin{align}
 I^{(2)}_2 &= {\rm Li}_2(1-a) + {a\over a-1}\log{a} - {\pi^2\over 6} \,. 
\end{align}
Note that there are no leading-power soft contributions corresponding to the full theory diagram in
Fig.~\ref{fig:scet1} involving the photon loop momentum $L^\mu$ in the numerator.

Using these integrals, 
an explicit evaluation of the diagram in Fig.~\ref{fig:scet1}a  yields
\begin{align}
  &\big( \sum |{\cal M}_{1}|^2 \big)_{{\rm Fig.\ref{fig:scet1}a,\,collinear}}
  \nonumber \\[5mm]
  \quad
  &= 2{\rm Re}\sum\Bigg(
    \parbox{25mm}{
      \begin{fmfgraph*}(70,40)
        \fmfset{arrow_len}{3mm}\fmfset{arrow_ang}{12}
        \fmfleftn{l}{3}
        \fmfrightn{r}{3}
        \fmfbottomn{b}{1}
        \fmf{phantom}{l3,x,v,y,r3}
        \fmf{photon, tension=0.6}{b1,v}
        \fmffreeze
        \fmf{fermion}{l3,v,y,r3}
        \fmf{photon, tag=1}{y,r2}
        \fmfipath{p[]}
        \fmfv{d.sh=circle,d.f=empty,d.si=.1w,l=$\times$,label.dist=0}{b1}
      \end{fmfgraph*}
    }
    +
    \parbox{25mm}{
      \begin{fmfgraph*}(70,40)
        \fmfset{arrow_len}{3mm}\fmfset{arrow_ang}{12}
        \fmfleftn{l}{3}
        \fmfrightn{r}{3}
        \fmfbottomn{b}{1}
        \fmftopn{t}{1}
        \fmf{phantom}{l3,x,v,y,r3}
        \fmf{photon, tension=0.6}{b1,v}
        \fmffreeze
        \fmf{fermion}{l3,x,v,r3}
        \fmf{photon}{x,t1}
        \fmfv{d.sh=circle,d.f=empty,d.si=.1w,l=$\times$,label.dist=0}{b1}
      \end{fmfgraph*}
    }
    \Bigg)^*
    \Bigg(
    \parbox{25mm}{
      \begin{fmfgraph*}(80,40)
        \fmfleftn{l}{4}
        \fmfrightn{r}{4}
        \fmfbottomn{b}{1}
        \fmf{phantom}{l4,x1,x2,v,y1,y2,r4}
        \fmf{photon, tension=0.6}{b1,v}
        \fmffreeze
        \fmf{fermion}{l4,x1,v,y1,y2,r4}
        \fmf{phantom,tag=1}{y1,r2}
        \fmfipath{p[]}
        \fmfiset{p1}{vpath1(__y1,__r2)}
        \fmfi{curly}{ point 0length(p1) of p1 -- point 3length(p1)/4 of p1}
        \fmf{curly,left=1}{v,y2}
        \fmf{plain,left=1}{v,y2}
        \fmfv{d.sh=circle,d.f=empty,d.si=.1w,l=$\times$,label.dist=0}{b1}
      \end{fmfgraph*}
    }
    +
    \parbox{25mm}{
      \begin{fmfgraph*}(70,40)
        \fmfset{arrow_len}{3mm}\fmfset{arrow_ang}{12}
        \fmfleftn{l}{3}
        \fmfrightn{r}{3}
        \fmfbottomn{b}{1}
        \fmftopn{t}{1}
        \fmf{phantom}{l3,x1,x2,,v,y1,y2,r3}
        \fmf{photon, tension=0.6}{b1,v}
        \fmffreeze
        \fmf{fermion}{l3,x1,x2,v,r3}
        \fmf{curly}{x2,t1}
        \fmf{curly,left=1}{v,x1}
        \fmf{plain,left=1}{v,x1}
        \fmfv{d.sh=circle,d.f=empty,d.si=.1w,l=$\times$,label.dist=0}{b1}
      \end{fmfgraph*}
    }  
    \Bigg) 
  \nonumber \\[5mm]
  \quad &= e^2 \sum  |{\cal M}_0|^2 {\alpha\over 4\pi} {2v\cdot v^\prime \over v\cdot k \,v^\prime\cdot k} {\rm Re}\bigg[ 4 I^{(1)}_1 + 2 I_2^{(2)} + (a\to b) \bigg]
  \nl
  \quad &= e^2 \sum |{\cal M}_0|^2 {\alpha\over 4\pi} {2v\cdot v^\prime \over v\cdot k \,v^\prime\cdot k}
     {\rm Re}\bigg[
     -2{\rm Li}_2(1-a) + {\pi^2\over 3} -{2a\over 1-a}\log{a}
     + (a\to b)
     \bigg]
\,.
\end{align}
Similarly, (extracting the overall factor
${\cal C} = e^2\sum|{\cal M}_0|^2 e_{\rm bare}^2 (4\pi)^{-2+\epsilon}\Gamma(1+\epsilon) m^{-2\epsilon}$,
and real part implied), 
\begin{align}
&{\cal C}^{-1}  \big( \sum |{\cal M}_{1}|^2 \big)_{{\rm Fig.\ref{fig:scet1}b,\,collinear}}
  = \left[ {1\over (v\cdot k)^2} + {1\over (v^\prime \cdot k)^2} -  {2v\cdot v^\prime \over v\cdot k\, v^\prime\cdot k} \right]
  \left[ -{4\over \epsilon^2} -{8\over \epsilon} \right]
  \nl
  &\qquad 
  + \left[ {1\over (v\cdot k)^2} + {1\over (v^\prime \cdot k)^2} \right]\left(  -16  \right)
  \nl
  &\qquad 
  + {2v\cdot v^\prime \over v\cdot k\, v^\prime\cdot k}
   {\rm Re}\bigg[
     2{\rm Li}_2(1-a) - {\pi^2\over 3} +{2a\over 1-a}\log{a} + 8
     + (a\to b)
     \bigg] \,,
  \nl
& {\cal C}^{-1}   \big( \sum |{\cal M}_{1}|^2 \big)_{{\rm Fig.\ref{fig:scet1}c,\,collinear}}
  = \left[ {1\over (v\cdot k)^2} + {1\over (v^\prime \cdot k)^2} -  {2v\cdot v^\prime \over v\cdot k\, v^\prime\cdot k} \right]
  \left( -{6\over \epsilon} \right) 
  \nl
  &\qquad
  + \left[ {1\over (v\cdot k)^2} + {1\over (v^\prime \cdot k)^2} \right](-8)
    + {2v\cdot v^\prime \over v\cdot k\, v^\prime\cdot k}
    \bigg[-{a(5a-4)\over (a-1)^2}\log{a} + {5a-4\over a-1} + (a\to b)  \bigg] \,,
\nl 
& {\cal C}^{-1} \big( \sum |{\cal M}_{1}|^2 \big)_{{\rm Fig.\ref{fig:scet1}d,\,collinear}}
 =    - {\cal C}^{-1}\big( \sum |{\cal M}_{1}|^2 \big)_{{\rm Fig.\ref{fig:scet1}c,\,collinear}} \,,
 \nl
 & {\cal C}^{-1}   \big( \sum |{\cal M}_{1}|^2 \big)_{{\rm Fig.\ref{fig:scet1}e,\,collinear}}
 =  \left[ {1\over (v\cdot k)^2} + {1\over (v^\prime \cdot k)^2}
   - {2v\cdot v^\prime \over v\cdot k\, v^\prime\cdot k}
   \right]\left( {6\over \epsilon}  +  8 \right) \,. 
\end{align}
Summing contributions, we find
\begin{align}
  \big( \sum |{\cal M}_{1}|^2 \big)_{{\rm collinear}}
&=   e^2\sum|{\cal M}_0|^2 {\alpha \over 4\pi}\left(m^2\over \mu^2\right)^{-\epsilon}
 \left[ {1\over (v\cdot k)^2} + {1\over (v^\prime \cdot k)^2}
   - {2v\cdot v^\prime \over v\cdot k\, v^\prime\cdot k}
   \right]
 \nl
 &\quad \times
 \bigg[
 -{4\over \epsilon^2} - {2\over\epsilon} -8 
    \bigg]
\end{align}
For the soft contributions,
\begin{align}
  {\cal C}^{-1}\big( \sum |{\cal M}_{1}|^2 \big)_{{\rm Fig.\ref{fig:scet1}a,\,soft}}
  &= \left[ {1\over (v\cdot k)^2} -  {v\cdot v^\prime \over v\cdot k\, v^\prime\cdot k} \right]
  \left[ -4 L^2 + 8\log{m a\over \lambda} L - {2\pi^2\over 3} \right] 
  \nl
  &\quad + \left[ {1\over (v^\prime \cdot k)^2} -  {v\cdot v^\prime \over v\cdot k\, v^\prime\cdot k} \right]
  \left[ -4 L^2 + 8\log{m b\over \lambda} L - {2\pi^2\over 3} \right] \,,
  \nl
  {\cal C}^{-1} \big( \sum |{\cal M}_{1}|^2 \big)_{{\rm Fig.\ref{fig:scet1}b,\,soft}}
  &= \left[ {1\over (v\cdot k)^2} -  {v\cdot v^\prime \over v\cdot k\, v^\prime\cdot k} \right]
  \left[ {4\over\epsilon} L - 8L \log{a} + 4 L^2 + {2\pi^2\over 3} \right]
  \nl
  &\quad
  +
  \left[ {1\over (v^\prime \cdot k)^2} -  {v\cdot v^\prime \over v\cdot k\, v^\prime\cdot k} \right]
  \left[ {4\over\epsilon} L - 8L \log{b}  + 4 L^2 + {2\pi^2\over 3} \right] \,,
  \nl
  {\cal C}^{-1} \big( \sum |{\cal M}_{1}|^2 \big)_{{\rm Fig.\ref{fig:scet1}c,\,soft}}
  &= \left[ {1\over (v\cdot k)^2} -  {v\cdot v^\prime \over v\cdot k\, v^\prime\cdot k} \right]
  \left[ {4\over\epsilon} - 8 \log{a} + 8\right]
  \nl
  &\quad
  +\left[ {1\over (v^\prime \cdot k)^2} -  {v\cdot v^\prime \over v\cdot k\, v^\prime\cdot k} \right]
  \left[ {4\over\epsilon} - 8 \log{b} + 8 \right] \,,
  \nl
  {\cal C}^{-1} \big( \sum |{\cal M}_{1}|^2 \big)_{{\rm Fig.\ref{fig:scet1}d,\,soft}}
  &= \left[ {1\over (v\cdot k)^2} -  {v\cdot v^\prime \over v\cdot k\, v^\prime\cdot k} \right]  
  \left[ -{4\over\epsilon} - 8 + 8 \log{a} \right]
  \nl
  &\quad
  +
  \left[ {1\over (v^\prime \cdot k)^2} -  {v\cdot v^\prime \over v\cdot k\, v^\prime\cdot k} \right]  
  \left[ -{4\over\epsilon} - 8 + 8 \log{b} \right] \,,
  \nl
  {\cal C}^{-1} \big( \sum |{\cal M}_{1}|^2 \big)_{{\rm Fig.\ref{fig:scet1}e,\,soft}}
  &= \left[ {1\over (v\cdot k)^2} + {1\over (v^\prime \cdot k)^2}  -  {2v\cdot v^\prime \over v\cdot k\, v^\prime\cdot k} \right]  
  \left[ -{4\over\epsilon} + 8 \log{\lambda\over m} \right] \,. 
\end{align}
Summing contributions,
\begin{align}
  \big( \sum |{\cal M}_{1}|^2 \big)_{{\rm soft}}
&=   e^2\sum|{\cal M}_0|^2 {\alpha \over 4\pi}\left(m^2\over \mu^2\right)^{-\epsilon}
 \left[ {1\over (v\cdot k)^2} + {1\over (v^\prime \cdot k)^2}
   - {2v\cdot v^\prime \over v\cdot k\, v^\prime\cdot k}
   \right]
 \nl
 &\quad \times
 \bigg[
 {1\over \epsilon}(4L-4) - 8(L-1)\log{\lambda\over m} 
    \bigg] \,. 
\end{align}
For the hard contribution, only Fig.~\ref{fig:scet1}b contributes, 
\begin{align}
  \big( \sum |{\cal M}_{1}|^2 \big)_{{\rm hard}}
&=   e^2\sum|{\cal M}_0|^2 {\alpha \over 4\pi}\left(m^2\over \mu^2\right)^{-\epsilon}
 \left[ {1\over (v\cdot k)^2} + {1\over (v^\prime \cdot k)^2}
   - {2v\cdot v^\prime \over v\cdot k\, v^\prime\cdot k}
   \right]
 \nl
 &\quad \times
 \bigg[
{4\over \epsilon^2} + {1\over \epsilon}(-4L + 6) + 2L^2 -6L  + 16  - {2\pi^2\over 3} 
    \bigg] \,. 
\end{align}

The contribution from the analog of Fig.~\ref{fig:scet1} with photon emitted 
from the initial state electron results in the same expressions with $a\leftrightarrow b$.
The sum of hard, collinear and soft contributions is identical at leading power  
to the full theory evaluation above.  

This analysis shows that 
individual diagrams contain nonvanishing contributions from soft photons emitted
interior to collinear photon loops.   As discussed around Eq.~(\ref{eq:Clead}),
the presence of multiple momentum modes contributing at leading power to the real-photon
phase space integration complicates a simple factorization argument.  Nonetheless,
an explicit evaluation reveals that factorization
holds in the sum over diagrams, at least through one loop order, consistent with
the direct evaluation (\ref{eq:direct}).  This leads to the simple expression
(\ref{eq:deltahat}), as required by the factorization formula (\ref{eq:facfull}).

\end{fmffile}

\end{document}